\documentclass{aa}  
\usepackage{graphicx}
\usepackage{txfonts}
\usepackage[pdfpagelabels=false]{hyperref}	% Hyperlinks
\hypersetup{colorlinks=true,linkcolor=blue,citecolor=blue,filecolor=blue,urlcolor=blue,}
\usepackage{longtable}
\usepackage{lscape}
\usepackage{xcolor}
\newcommand{\Msun}{{M}_{\odot}}

\usepackage{natbib}
\defcitealias{ChruslinskaNelemans19}{ChN19}
\defcitealias{Chruslinska24_OFe}{Ch24}
\defcitealias{Chruslinska21}{Ch21}
\defcitealias{Popesso23}{P23}
\begin{document}

   \title{Trading oxygen for iron}
   \subtitle{II. Oxygen- versus iron-dependent cosmic star formation history}
\titlerunning{cosmic evolution of O \& Fe abundances}

   \author{M. Chru{\'s}li{\'n}ska
          \inst{1,2}, 
M. Curti\inst{1},
%\orcidlink{0000-0002-2678-2560}
    R. Pakmor\inst{2},
    %\orcidlink{0000-0003-3308-2420}
    A. De Cia\inst{1},
    %\orcidlink{0000-0003-2082-1626}
    J. Matthee\inst{3},
    %\orcidlink{}
    A. Bhagwat\inst{2},
    %\orcidline{0000-0003-0275-5506}
    S. Monty\inst{4, 5}
    %\orcidlink{0000-0002-9225-5822}
          }
  \institute{European Southern Observatory, Karl-Schwarzschild-Str. 2, 85748 Garching, Germany 
\and Max Planck Institute for Astrophysics, Karl-Schwarzschild-Str. 1, D-85748 Garching, Germany
\and Institute of Science and Technology Austria, Am Campus 1, 3400 Klosterneuburg, Austria
\and Center for Interdisciplinary Exploration and Research in Astrophysics (CIERA), Northwestern University, 1800 Sherman Avenue, Evanston, IL 60201, USA
\and Department of Astronomy, New Mexico State University, Las Cruces, NM 88003, USA\\
             }

   \date{Received November , 2025; accepted XX XX, 2025}

  \abstract{
Due to their different nucleosynthetic origin, a stellar population produces oxygen (O) and iron (Fe) on different timescales and their relative abundance can deviate strongly from solar. Galaxy formation models should treat these elements separately, as they play a distinct role in shaping physical phenomena. For example, oxygen mainly sets the gas cooling rate, while the iron abundance sets stellar atmosphere opacities impacting stellar evolution, spectra and feedback. Observations of star-forming galaxies usually only constrain gas-phase oxygen abundance, vastly limiting our capabilities of separating the cosmic evolution of oxygen and iron. Here, we present an observationally-motivated framework to scale the cosmic evolution of O and Fe abundances. We apply the relation between the $\alpha$-enhancement and galaxies’ specific star formation rate ([O/Fe]–sSFR) to derive the Fe and O-dependent cosmic star-formation history (cSFH). We find that star formation with near-solar O/Fe is rare: at least $70\%$ of the integrated cosmic stellar mass forms at non-solar O/Fe. The cosmic average metallicity is generally lower in [Fe/H] than in [O/H] (by up to a factor 3), with the offset increasing from $z=0$ to redshifts $z\approx 3$ and then approaching the core-collapse O/Fe ratio. We validate our results against samples that probe the Fe-dependent cSFH in different regimes such as absorption-derived $\langle{\rm [Fe/H]}\rangle$ from long gamma-ray bursts. Our results impact the interpretations of stellar and galaxy spectra and the predicted rates of transients, especially those linked to metal-poor progenitors (e.g., black hole mergers).
  }

   \maketitle

\nolinenumbers
\section{Introduction}
The chemical composition of material within galaxies evolves as nucleosynthetic products from stars are ejected and mixed into the surrounding medium. Increasing metallicity (abundance of elements heavier than helium)
alters the physics and observable properties of both stars \citep{Garcia21,Eldridge22,Vink22} and galaxies \citep{Bromm03,MaiolinoMannucci19,Kewley19}, but not all elements affect evolution equally. 
Among them, oxygen and iron provide a striking example: they regulate different key processes and their abundances evolve on different timescales. Oxygen largely sets gas cooling, regulating how efficiently gas is converted into stars \citep{Bromm03,Richings_2014,katz2022,Sharda23}; iron-group elements control stellar opacities and radiation-driven winds \citep{Iglesias96,Vink01}, shaping stellar lives and deaths \citep{Puls08,Langer12,Smith14,Maynet15,Heger23}, transient formation and properties \citep{LangerNorman06,Belczynski10,Perley16,Schulze18,Chruslinska24}, ionizing-radiation output \citep[e.g.][]{Leitherer99,StanwayEldridge18,Gotberg19,Gotberg20}, and feedback on galactic scales \citep{Eldridge22}.
Oxygen is produced by sources linked to massive stars and released promptly after star formation \citep{Woosley02,Heger03,Janka07,Schneider21,KobayashiTaylor23}. Iron additionally has a substantial contribution from Type~Ia supernovae \citep{Tinsley79,Nomoto97,MaozMannucciNelemans14,Kobayashi20} with a wide delay-time distribution \citep{Nomoto82,IbenTutukov84,Webbink84,Greggio05,MaozMannucci12,WangHan12,MaozMannucciNelemans14,LivioMazzali18}. Consequently, their relative abundance can depart from the solar ratio by a factor of $\gtrsim$ 5 \citep{MatteucciGreggio86,Wheeler89,KobayashiTaylor23,Chruslinska24_OFe,Monty25}. 
Observations of star-forming galaxies typically constrain only oxygen, which is the most abundant metal with strong emission lines \citep{MaiolinoMannucci19} available for growing samples across redshifts \citep{Nakajima23,Chemerynska24,Curti24,Revalski24,Sanders24}. By contrast, the iron abundance in star-forming gas is particularly difficult to determine (e.g. due to faint lines and strong dust depletion) and will likely remain unavailable for representative samples in the near term \citep{Chruslinska24_OFe}.
Consequently, oxygen abundance is often used as a proxy for it, assuming a scaling that maintains relative abundance ratios as determined for the Sun. This can lead to important and as yet unaccounted for errors - here we propose a method to address this issue.
\\
In Paper I, \cite{Chruslinska24_OFe} (\citetalias{Chruslinska24_OFe}) introduced the relation between oxygen-to-iron abundance ratio and the specific star formation rate of galaxies ([O/Fe] - sSFR relation) as an effective to translate oxygen to iron abundances.  
\citetalias{Chruslinska24_OFe}
showed that the relation is supported by cosmological simulations  \citep{Schaye15,Crain15,Weinberger17,Pillepich18} and basic theoretical expectations \citep[see also][]{MattheeSchaye18,Kashino22}, and 
 used a homogenised sample of star-forming galaxies to provide empirical constraints. Here we apply this relation to infer the iron-dependent cosmic star formation history, $f_{\rm SFR}(\mathrm{Z_{Fe/H}},t)$, and compare it with its oxygen-based counterpart, $f_{\rm SFR}(\mathrm{Z_{O/H}},t)$. To this end, we expand and update the framework developed in \citet{ChruslinskaNelemans19} (\citetalias{ChruslinskaNelemans19}) and \citet{Chruslinska21} (\citetalias{Chruslinska21}), which estimates $f_{\rm SFR}(\mathrm{Z_{O/H}},t)$ by combining empirical distributions of galaxy properties (star formation rate SFR, gas-phase oxygen abundance $Z_{O/H}$, stellar mass $M_{*}$) with number statistics, while propagating the associated uncertainties.
 As a verification step, we compare our $f_{\rm SFR}(\mathrm{Z_{Fe/H}},t)$ predictions with samples of Milky Way globular clusters \citep{Carretta09,Vandenberg13} and RR~Lyrae stars in local dwarfs \citep{Bellazzini25}, which probe different parts of the cosmic star formation rate (SFRD) distribution in iron (Sec. \ref{sec: Fe-cross-check-samples}). 
In Sec. \ref{sec: Fe/H avg. comparison} we further compare our results with the absorption-based cosmic SFRD-weighted average metallicity probed by long gamma-ray bursts \citep[LGRBs;][]{Heintz23}.
\\
Where appropriate, we adopt a standard flat cosmology with $\Omega_{\rm M}$ = 0.3, $\Omega_{\rm \Lambda}$ = 0.7, and $H_{0}$ = 70 km s$^{-1}$ Mpc$^{-1}$ \citep{Hinshaw13} and assume a \citet{Kroupa01} initial mass function (IMF).
The terms 'redshift' ($z$), 'cosmic time' ($t$) and 'lookback time' are used interchangeably throughout the paper ($z$= lookback time = 0 at present day), the relation between them is fixed within the assumed cosmological model.
We use the solar reference abundances from \citet{GrevesseSauval98} for consistency with previous work (12+log$_{10}$(O/H)$_{\odot}$=8.83, 12+log$_{10}$(Fe/H)$_{\odot}$=7.5, log$_{10}$(O/Fe)$_{\odot}$=1.33). Where necessary, we shift abundances of literature samples to this solar reference scale.
Different solar reference values are reported and used across the literature \citep[especially for oxygen different estimates differ by $\sim$0.15dex, e.g.][]{Lodders09,Asplund09,Asplund21}.
This choice only affects the numerical results reported for abundances relative to solar. Where these results cannot be straightforwardly converted to a different solar reference, we comment on the impact this has.
We use the following notation to distinguish absolute abundances: $Z_{X/H}$=12+log$_{10}$(X/H) and abundances relative to solar: [X/H] = $Z_{X/H}$ - $Z_{X/H; \odot}$, where the index 'X=O' for oxygen and 'X=Fe' for iron, respectively.

\section{Method: trading oxygen for iron using the [O/Fe] - sSFR relation of galaxies}\label{sec: method}

\begin{figure}[htbp!]
    \centering
    \includegraphics[width=0.5\textwidth]{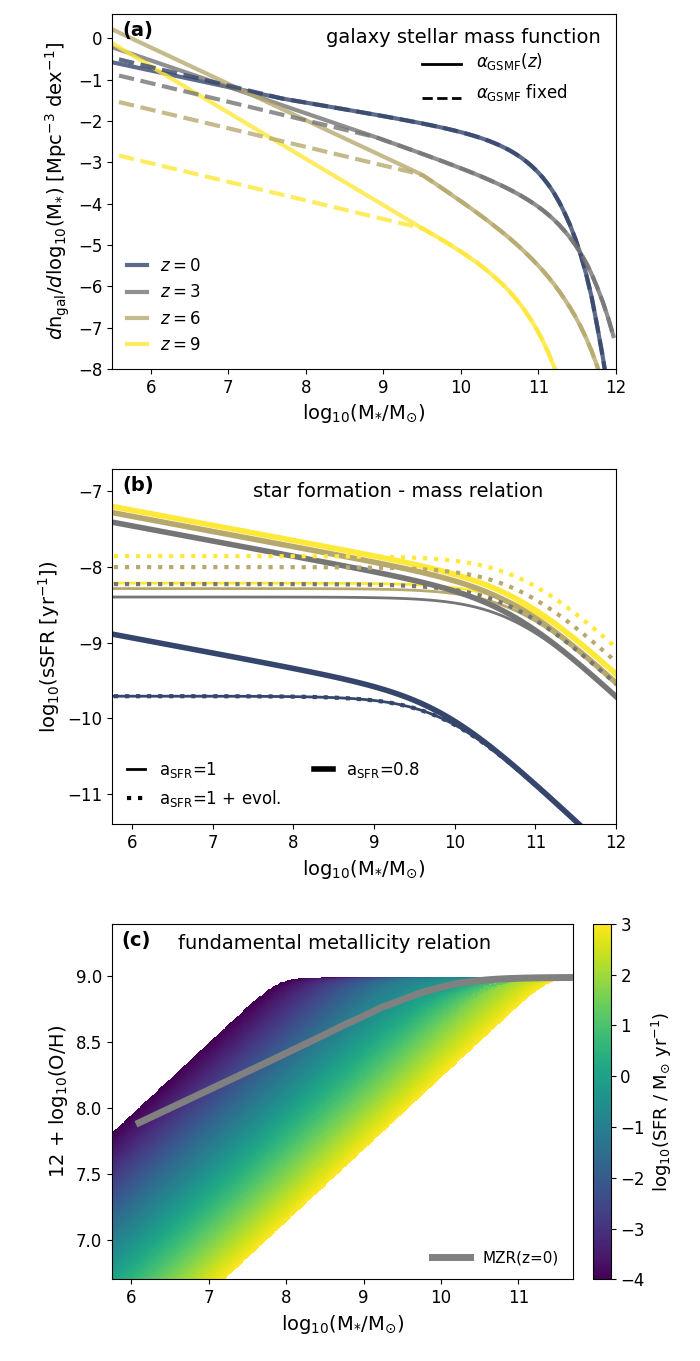}
    
    \caption{
    Galaxy stellar mass function (top panel) which provides the number density of galaxies as a function of $M_{*}$, and scaling relations linking the galaxy properties used in our framework: the sSFR–$M_{*}$ relation (derived from the \citetalias{Popesso23} SFMR; middle panel), the joint dependence of $Z_{O/H}$, SFR, and $M_{*}$ shown as a 2D projection of the redshift-invariant FMR (bottom panel, also showing the $z\sim0$ MZR). Colours in the top and middle panels indicate different redshifts, illustrating the evolution of these relations. 
Different line styles in the top and middle panels show considered variations in the assumed low-$M_{*}$ slope and redshift evolution of the relations (see Appendix~\ref{app: variations}).
The evolution of the $Z_{O/H}$ scaling relations is shown in Fig. \ref{fig: oxygen-MZR}.
    }
    \label{fig: scaling rel.}
\end{figure}

We aim to self-consistently derive and compare the oxygen-based f$_{\rm SFR}$(Z$_{O/H}$,$t$) and iron-based f$_{\rm SFR}$(Z$_{Fe/H}$,$t$). 
We follow the phenomenological approach detailed in \citetalias{ChruslinskaNelemans19} and \citetalias{Chruslinska21}. 
The key additional step introduced in this work is the application of the [O/Fe] -- sSFR relation of galaxies from \citetalias{Chruslinska24_OFe}, which allows us to estimate f$_{\rm SFR}$(Z$_{Fe/H}$,t).
The core idea of our framework is summarized below, while further details can be found in the appendix \ref{app: method}.
Our approach requires knowledge of the following main ingredients:
\begin{itemize}
\item[1)] [O/Fe] - specific SFR relation (redshift invariant)
\item[2)] Distribution of sSFR of galaxies as a function of redshift 
\item[3)] (a) Distribution of Z$_{O/H}$ of galaxies \\
(b) Relation linking galaxy Z$_{O/H}$ and sSFR \\
(c) Evolution of (3a) and potentially (3b) with redshift
\item[4)] Number density of galaxies $n_{gal}$ with different sSFR and Z$_{O/H}$ as a function of redshift 
\end{itemize}
These are illustrated in Figure \ref{fig: scaling rel.}  (2-4), Figure \ref{fig: OFe-sSFR rel.} (1) and Figure \ref{fig: oxygen-MZR} (3c).
In \citetalias{ChruslinskaNelemans19,Chruslinska21},
(2-4) are determined based on a compilation of observational studies and extrapolated where necessary. Distributions (2) and (3a) are characterised as a function of the galaxy's stellar mass $M_{*}$. 
At a fixed $z$, galaxies follow a ridge in the Z$_{O/H}$ - log$_{10}(M_{*})$ plane with characteristic scatter $\sigma_{MZR}$ in Z$_{O/H}$, referred to as the mass-metallicity relation \citep[MZR, gray line in the bottom panel in Fig. \ref{fig: scaling rel.}, e.g.][see the review by \citealt{MaiolinoMannucci19}]{Tremonti04,Maiolino08,Curti20,Sanders20, Nakajima23,Curti24,Scholte24}.
We obtain the distribution of sSFR (middle panel in Fig. \ref{fig: scaling rel.}) starting from the distribution of galaxies observed in the log$_{10}(SFR)$ - log$_{10}(M_{*})$ plane. The main ridge in this plane defines the galaxy main sequence \citep[e.g.][here called the star formation-mass relation -- SFMR, for consistency with earlier work]{Brinchmann04,Speagle14,Tomczak16,Popesso23,Simmonds25,DiCesare25} with characteristic scatter $\sigma_{SFR}$=0.3 dex. 
A bimodality in the log$_{10}(SFR)$ - log$_{10}(M_{*})$ plane is sometimes observed, with the secondary ridge occupied by starburst (SB) galaxies and referred to as the SB sequence \citep[e.g.][]{Rodighiero11,Sargent12,Schreiber15,Caputi17,Rinaldi25}. 
If Z$_{O/H}$ and SFR (sSFR) were not correlated, these galaxy properties could be easily connected via $M_{*}$. 
In general this is not the case, and additional relation linking Z$_{O/H}$, SFR and $M_{*}$ (3b) takes into account the observed anticorrelation between the galaxy's Z$_{O/H}$ and sSFR \citep[called fundamental metallicity relation - FMR,  bottom panel in Fig. \ref{fig: scaling rel.}e.g.][]{Ellison08,Mannucci10,Mannucci11,Yates12,AndrewsMartini13,Lara-Lopez13,Sanders18,Curti20}.
 FMR is found to be $z$-invariant in the models/simulations  (\citealt{Dave11,Lagos16,DeRossi17,Torrey18} but see \citealt{Garcia24}) and is consistent with no evolution according to observations \citep[e.g.][]{Cresci19,Sanders20} at least up to z$\sim$3.  If the FMR is $z$-invariant, (3c) the evolution of galaxy $Z_{O/H}$ with $z$ is determined by the evolution of other galaxy properties.
However, recent works obtain conflicting results regarding the FMR and its universality in the early Universe
\cite[e.g.][identify z$\gtrsim$6 galaxies showing significant departures from the z=0 FMR]{Heintz23,Langeroodi23, 
Nakajima23,Curti23,Curti24} and at low $M_{*}\lesssim10^{9}\Msun$ \citep{Kotiwale25,Laseter25}.
We note that $Z_{O/H}$ at z$\gtrsim$3 (and 3c) were essentially unconstrained prior to JWST, and  the works of \citetalias{ChruslinskaNelemans19,Chruslinska21} rely on different extrapolations in this regime.
In Section \ref{sec: oxygen updates} we review our assumptions in light of these new observational results and rule out some of those extrapolations, making a crucial update to our framework.
\\
Once (2) and (3) are fixed, the Z$_{Fe/H}$ of each galaxy (characterised by $z$, $M_{*}$, Z$_{O/H}$ and SFR) can be estimated using (1) (see Sec. \ref{sec: which O/Fe-sSFR} and Fig. \ref{fig: OFe-sSFR rel.}).
Following \citetalias{ChruslinskaNelemans19}, to account for the fact that star formation within a galaxy occurs over a range of metallicities, we distribute the SFR in metallicity as a Gaussian centred on the $Z_{\rm O/H}$ and $Z_{\rm Fe/H}$ values assigned from the scaling relations. For simplicity, we adopt a common dispersion, $\sigma_{\nabla \rm O/H}=\sigma_{\nabla \rm Fe/H}=0.14$~ dex, with the value guided by the observed $Z_{\rm O/H}$ distribution in HII–regions of $z\sim0$ disk galaxies \citep[][see appendix \ref{app: scatter} for additional discussion]{Sanchez-Menguiano16,Sanchez-Menguiano17}.
The role of step (4) is to provide the `weighting factor', i.e. account for the fact that galaxies with different properties are not equally abundant in the cosmic volume when counting their contribution to the cosmic star formation budget.
All relevant galaxy properties in our framework can be linked via $M_{*}$, and the weighting $n_{gal}$($M_{*}$,$z$) is obtained from the galaxy stellar mass function \citep[GSMF, top panel in Fig. \ref{fig: scaling rel.}, e.g.][]{Baldry12,Muzzin13,Stefanon21}.
The product of (2) and (4) integrated over the full relevant range of $M_{*}$ ($10^{6}-10^{12} \Msun$ in this work) gives the total (i.e. at all metallicities) star formation rate density (SFRD). 
The result of this integration as a function of $z$ gives the cosmic star formation history (cSFH, section \ref{sec: cSFH}).
By summing only the contribution of galaxies with a given Z$_{O/H}$ or Z$_{Fe/H}$, we derive the SFRD as a function of metallicity, i.e. either oxygen- or iron-based metallicity-dependent cSFH. See appendix \ref{app: method} for the derivation.
\\\newline
For our initial estimate we assume:\\
(1) [O/Fe] - sSFR relation corresponding to the "mixed" Fe enrichment variation defined in Sec. \ref{sec: which O/Fe-sSFR}
\\
(2) SFMR from \cite{Popesso23} (\citetalias{Popesso23}), based on a homogenized set of observational literature estimates spanning $0<z<6$, but with additional evolution in the normalisation at high $z$ (dotted lines in the middle panel in Fig. \ref{fig: scaling rel.}). This modification is motivated by the recent results of \cite{Simmonds25}, constraining the SFMR up to $z\sim9$. 
Table \ref{tab: variations} (row \emph{"$a_{\rm SFR}$=1+ evol."}) provides implementation details.
We include a small fraction of SB ($f_{SB}$=3\%, as in  \citealt{Boco21},\citetalias{Chruslinska21}), which yields negligible SB contribution to the cSFH (see appendix \ref{sec, app: SB} for further details).
\\
(3a) MZR($z=0$) matching \citet{Curti20} with the asymptotic value at high $M_{*}$ ($Z_{O/H; \rm MZR0}$)  shifted by +0.2 dex (gray line in the bottom panel in Fig. \ref{fig: scaling rel.}, see Tab. \ref{tab: variations} for details). \citetalias{Chruslinska24_OFe} show that such systematic shift in $Z_{O/H}$ estimated from electron temperature-sensitive auroral lines (and the derived strong line calibrations, including \citealt{Curti20}) may be necessary to reconcile the average [O/Fe] of young star-forming galaxies and old Milky Way (MW) stars.
\\
(3b) FMR as described in \citetalias{Chruslinska21} (colour-mapped in the bottom panel in Fig. \ref{fig: scaling rel.}), determined by MZR($z=0$), SFMR($z=0$) and a parameter $\nabla_{\rm FMR0}$, characterising the strength of the Z$_{O/H}$ - SFR correlation at high SFR and low/intermediate $M_{*}$\footnote{$\delta \rm MZR$ = -$\nabla_{\rm FMR0} \cdot \delta \rm SFMR$, where $\delta \rm MZR$ and $\delta \rm SFMR$ are the galaxy’s offsets from the respective relations.} (see appendix \ref{app: FMR} for further details). We use $\nabla_{\rm FMR0}=0.27$ as in \citetalias{Chruslinska21}.
\\
(3c) Fixed FMR at $z\leqslant$3,  and the FMR with decreasing normalisation at $z>$3 as discussed in Sec \ref{sec: oxygen updates}. This takes into account the potential additional Z$_{O/H}$ evolution beyond that resulting from a $z$-invariant FMR, as allowed by current observational constraints.
\\
(4) GSMF from \citetalias{ChruslinskaNelemans19} extrapolated to low $M_{*}$=10$^{6}\Msun$ with a $z$-dependent slope $\alpha_{\rm GSMF}(z)$ and updated to include more recent results (solid lines in top panel in Fig. \ref{fig: scaling rel.}), as described in Appendix \ref{sec, app: GSMF}.
\\
We refer to the model characterised by the above choices (1–4) as the 'example variation'.
We use it in Section \ref{sec: results-example variation} to discuss the main characteristics of our results and show it in more detail.
In Sec. \ref{sec: solar O/Fe} and beyond, we also explore alternative assumptions to assess the sensitivity of our results to these choices. The variations, summarized in Table \ref{tab: variations} in Appendix \ref{app: variations}, are selected to capture the major uncertainties in our framework while remaining broadly consistent with current constraints on the  [O/Fe] -- sSFR relation (Sec. \ref{sec: which O/Fe-sSFR}), MZR (Sec. \ref{sec: oxygen updates}), and cSFH (Sec. \ref{sec: cSFH}).

\subsection{The relation between [O/Fe] - specific SFR}\label{sec: which O/Fe-sSFR}
\begin{figure}[htbp]
    \centering
    \includegraphics[width=0.5\textwidth]{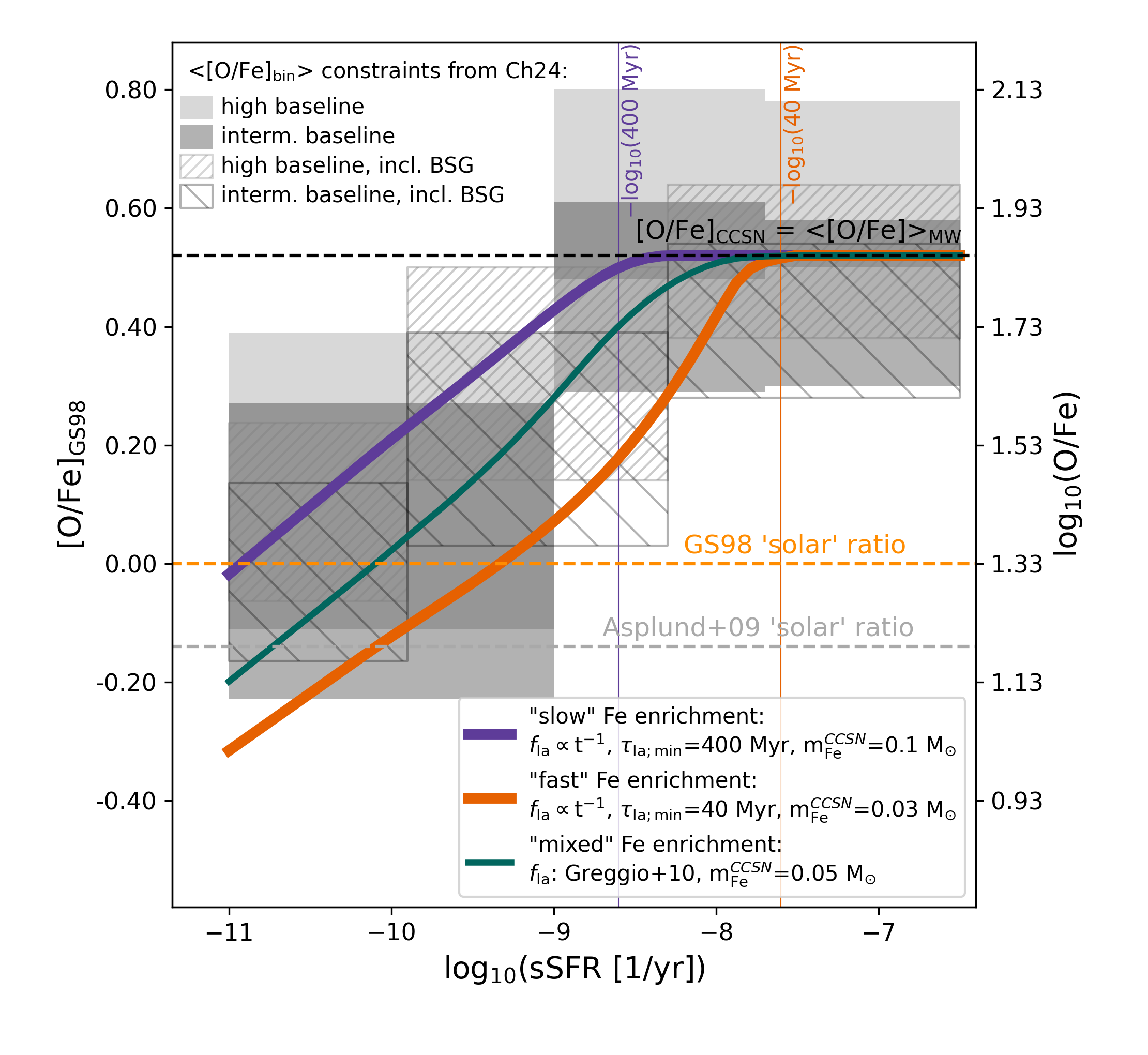}
    
    \caption{
The [O/Fe] - sSFR relation of galaxies.
coloured lines show three model variations considered in this paper (purple - "slow" Fe enrichment, orange - "fast" Fe enrichment and green - "mixed" Fe enrichment), which span the range allowed by observational constraints discussed in \citetalias{Chruslinska24_OFe} and shown as solid/hatched areas (see text for the details).
The horizontal black line indicates the average abundance ratio of the Milky Way thick disc/halo dwarf stars with [Fe/H] $<$–2 from \cite{Amarsi19}. The horizontal orange line indicates our reference solar log$_{10}$(O/Fe)$|_{\odot}$=1.33 dex \citep{GrevesseSauval98}. The zero point on the \cite{Asplund09} solar scale is shown as a gray dashed line.
    }
    \label{fig: OFe-sSFR rel.}
\end{figure}
Figure \ref{fig: OFe-sSFR rel.} summarizes constraints on the [O/Fe] - sSFR relation derived by \citetalias{Chruslinska24_OFe}, who also show that most of current theoretical expectations are consistent with them. 
These constraints were obtained after homogenising the $Z_{\mathrm{O/H}}$ and $Z_{\mathrm{Fe/H}}$ measurements of star-forming galaxies, which were compiled from the literature and binned in sSFR. 
They are illustrated as gray boxes in Fig. \ref{fig: OFe-sSFR rel.}, which show the [O/Fe] in three log$_{10}$(sSFR) bins and span 0.13 - 99.87 percentiles of the mean [O/Fe] in each bin.
\citetalias{Chruslinska24_OFe} homogenize oxygen and iron abundances by correcting for known average systematic offsets (putting them on a 'common baseline') and excluding cases that cannot be consistently treated. All abundances are first converted to a common solar reference scale \citep{GrevesseSauval98}, removing offsets of up to $\gtrsim$0.14 dex arising from differences in the adopted solar abundances.
For oxygen, only H II-region-based estimates are used, relying either on temperature-sensitive emission lines or on empirically calibrated strong-line diagnostics tied to such measurements. Some literature values include additional systematic shifts, in particular corrections of $\approx$0.2$-$0.24 dex to place abundances on the recombination-line scale, or dust depletion corrections of typically $\approx$0.1 dex. \citetalias{Chruslinska24_OFe} either remove these corrections or apply them uniformly to all data, depending on the chosen common baseline.
For iron, the methods used in the literature are less uniform and the associated systematics are less well quantified than for oxygen. \citetalias{Chruslinska24_OFe} restrict their analysis to $Z_{\mathrm{Fe/H}}$ estimates derived either from spectra of young stars or from rest-frame UV spectra of galaxies. The latter depend on the adopted stellar population synthesis (SPS) models, with an average difference of $\approx0.1$ dex between values derived using Starburst99 \citep{Leitherer99} and BPASS \citep[][i.e. the two most commonly used SPS libraries in the compiled datasets]{Stanway16,StanwayEldridge18}. 
All of these systematics are accounted for using either the average offsets listed above or, where available, the exact values reported in the original studies (see Table E.1 in \citetalias{Chruslinska24_OFe}).
\citetalias{Chruslinska24_OFe} define several common baseline choices. In Fig. \ref{fig: OFe-sSFR rel.}, we only show those for which the high sSFR 'plateau' of the [O/Fe]--sSFR relation is at the value consistent with the average <[O/Fe]>$_{\rm MW}$ of old metal-poor MW stars. These values are expected to be similar, reflecting the integrated yields of massive, metal-poor stars and the associated core-collapse supernovae (CCSNe; that eject material with supersolar O/Fe) prior to significant enrichment from delayed sources.
Hatched boxes show (tighter) constraints derived assuming that blue supergiant (BSG)-based metallicity \citep{Bresolin16,Bresolin22} can be used as a measure of galaxy's Z$_{Fe/H}$.
The coloured lines in Fig. \ref{fig: OFe-sSFR rel.} show three model variations of the [O/Fe] - sSFR relation that span these constraints and which we consider here. These relations are computed following eq. 1 from \citetalias{Chruslinska24_OFe}, recalled below: 
\begin{multline}\label{eq: OFe}
\rm [O/Fe] = [O/Fe]_{CCSN} \\
\rm - log_{10}\left( 1 + \frac{m^{Ia}_{Fe}}{m^{CCSN}_{Fe}} \frac{N_{Ia0}}{k_{CCSN}} \frac{\int_{0}^{t} SFR(t`) f_{Ia}(t-t`) \,dt`}{SFR} \right)
\end{multline}
where $[\mathrm{O/Fe}]_{\mathrm{CCSN}}$
sets the oxygen-to-iron abundance ratio relative to solar in high sSFR regime dominated by CCSN enrichment, $m^{CCSN}_{Fe}$ and $m^{Ia}_{Fe}$ are the average iron masses ejected per CCSN and SN~Ia event, $k_{CCSN}$ is the CCSN formation efficiency (number of CCSNe per unit stellar mass formed),$\rm N_{Ia0}$ is the SN Ia formation efficiency (i.e. the Hubble time-integrated number of SN Ia formed per unit stellar mass) , and $f_{Ia}$ is the SN~Ia delay-time distribution (DTD), normalised to unity over the Hubble time.
We stress that these parameters should be understood as effective quantities and that their direct interpretation in terms of supernova yields or specific enrichment channels is not straightforward. Eq.~\ref{eq: OFe} simply provides a convenient parametrization that captures the main features of the relation and preserves its connection to delayed iron enrichment relative to oxygen  - the primary physical driver.
\\
In Paper I, \citetalias{Chruslinska24_OFe} demonstrate using cosmological simulations of galaxy evolution (EAGLE \citealt{Crain15,Schaye15,McAlpine16} and TNG \citealt{Weinberger17,Pillepich18,Nelson19_TNG}) that the average $[\mathrm{O/Fe}]$--sSFR relation can be reproduced with Eq.~\ref{eq: OFe} using fixed effective values of the iron yields ($m^{CCSN}_{Fe}$, $m^{Ia}_{Fe}$) and the average abundance $[\mathrm{O/Fe}]_{\mathrm{CCSN}}$ (see Sec. 2.5 and Fig. 4 therein). This holds despite the fact that these simulations adopt different metallicity-dependent yield tables and include additional processes (e.g. galaxy mergers and gas flows) that can affect chemical evolution of individual galaxies. 
In these examples, the assumed metallicity-dependent yields and CCSN progenitor mass range affect mostly the effective normalisation $[\mathrm{O/Fe}]_{\mathrm{CCSN}}$. 
 \citetalias{Chruslinska24_OFe} identify SN Ia DTD and the relative iron production rate in SN Ia compared to CCSNe,
$C_{\rm Ia/CC} \coloneqq \frac{m^{Ia}_{Fe}}{m^{CCSN}_{Fe}} \frac{N_{Ia0}}{k_{CCSN}},
$, as the main factors controlling the shape of the relation. These parameters can be understood more broadly as the delay in enhanced iron production (whether arising from SN~Ia, low-mass CCSNe, or both) and the relative efficiency of prompt and delayed iron and oxygen enrichment channels.
As the underlying nucleosynthetic processes do not depend on redshift, the average [O/Fe] -- sSFR relation is not expected to evolve. However, galaxy populations at different redshifts are expected to occupy different regions of the $[\mathrm{O/Fe}]$–sSFR plane: at higher redshifts, galaxies are typically younger and have higher sSFR (see middle panel in Fig. \ref{fig: scaling rel.}), with chemical enrichment dominated by massive-star progenitors, and thus exhibit, on average, higher $[\mathrm{O/Fe}]$ ratios than galaxies at lower redshift.
\\
Following \citetalias{Chruslinska24_OFe}, we fix $\rm [O/Fe]_{CCSN}$=<[O/Fe]>$_{\rm MW}$=0.52 dex\footnote{Our results in terms of Z$_{Fe/H}$ and O/Fe values can simply be shifted by a constant to match other choices} to match average abundance ratio of metal-poor MW stars from \citealt{Amarsi19} (used as a baseline for their careful treatment of non-LTE effects).
The flattening at $\rm [O/Fe]_{CCSN}$ at the high-sSFR end of the relation is analogous to that found for Milky Way stars at low [Fe/H] in conventional Tinsley--Wallerstein diagrams (showing stellar [Fe/H] versus their enhancement in $\alpha$-elements, \citealp{Wallerstein62,Tinsley79}), 
however, the very low [Fe/H]$\lesssim$-2.5 regime is poorly constrained by observations. 
Detailed galaxy chemical evolution models often predict a gradual change of slope at high $\rm [O/Fe]$ \citep[e.g.][]{Matteucci21}  rather than a strict plateau (as in eq. \ref{eq: OFe}). In this view, our results may underestimate [O/Fe] and the low-Fe tail of the $f_{SFR}(Z_{Fe/H},t)$ at high redshift.
On the other hand, rare transients associated with very massive, metal-poor stars (e.g. pair-instability supernovae \citealt{HegerWoosley02}, or certain hypernova scenarios \citealt{Grimmet21}) can produce ejecta with significantly lower $[\mathrm{O/Fe}]$ than standard CCSNe. If such events occur with non-negligible frequency, they would introduce additional scatter and potentially a population of outliers below the high-sSFR plateau that is not accounted for here. Further discussion of this point can be found in sections 4.2 and 6 in \citetalias{Chruslinska24_OFe}.
\\
For the purpose of this work, we assume a power-law SN Ia DTD ($\rm f_{Ia}\propto t^{-\alpha_{Ia}}$ at t$>\tau_{\rm Ia; min}$ and zero at t$<\tau_{\rm Ia; min}$) with $\alpha_{Ia}=1$. This particular choice does not affect our conclusions (other $\rm f_{Ia}$ parametrisations are possible, but they degenerate into [O/Fe]-sSFR relations, see examples in \citetalias{Chruslinska24_OFe}). What is important, however, is the uncertain turning point of the relation, which for the assumed DTD is set by the minimum SN Ia delay $\tau_{\rm Ia; min}$ (thin vertical lines in Fig. \ref{fig: OFe-sSFR rel.}).
Therefore, we define two variations:
\begin{itemize}
    \item[(i)] `Fast' Fe enrichment: $\tau_{\rm Ia; min}$=40 Myr and $\rm m^{CCSN}_{Fe}$=0.03 $\Msun$ (or C$_{\rm Ia/CC}$=2.5)
    \item[(ii)] `Slow' Fe enrichment: $\tau_{\rm Ia; min}$=400 Myr and $\rm m^{CCSN}_{Fe}$=0.1 $\Msun$ (or C$_{\rm Ia/CC}$=0.74)
\end{itemize}
Those choices can be considered extreme based on current theoretical expectations (see sec. 5 in \citetalias{Chruslinska24_OFe} and references therein) and until tighter constraints are available, we use (i) and (ii) to bracket the uncertainty in the [O/Fe]-sSFR relation. 
Additionally, we consider an intermediate scenario:\\
(iii) `Mixed' Fe enrichment:  $\rm f_{Ia}$ from \citet{Kashino22} and $\rm m^{CCSN}_{Fe}$=0.05 $\Msun$ (or C$_{\rm Ia/CC}$=1.48)\\
where the $\rm f_{Ia}$ is based on an analytic theoretical model from \citet{Greggio05,Greggio10} 
 which allows for a mixed contribution of different proposed SN Ia progenitors.

\subsection{Z$_{O/H}$ at high $z$: our model in light of JWST} \label{sec: oxygen updates}

\begin{figure}[h!]
\centering
 \includegraphics[width=1\columnwidth]{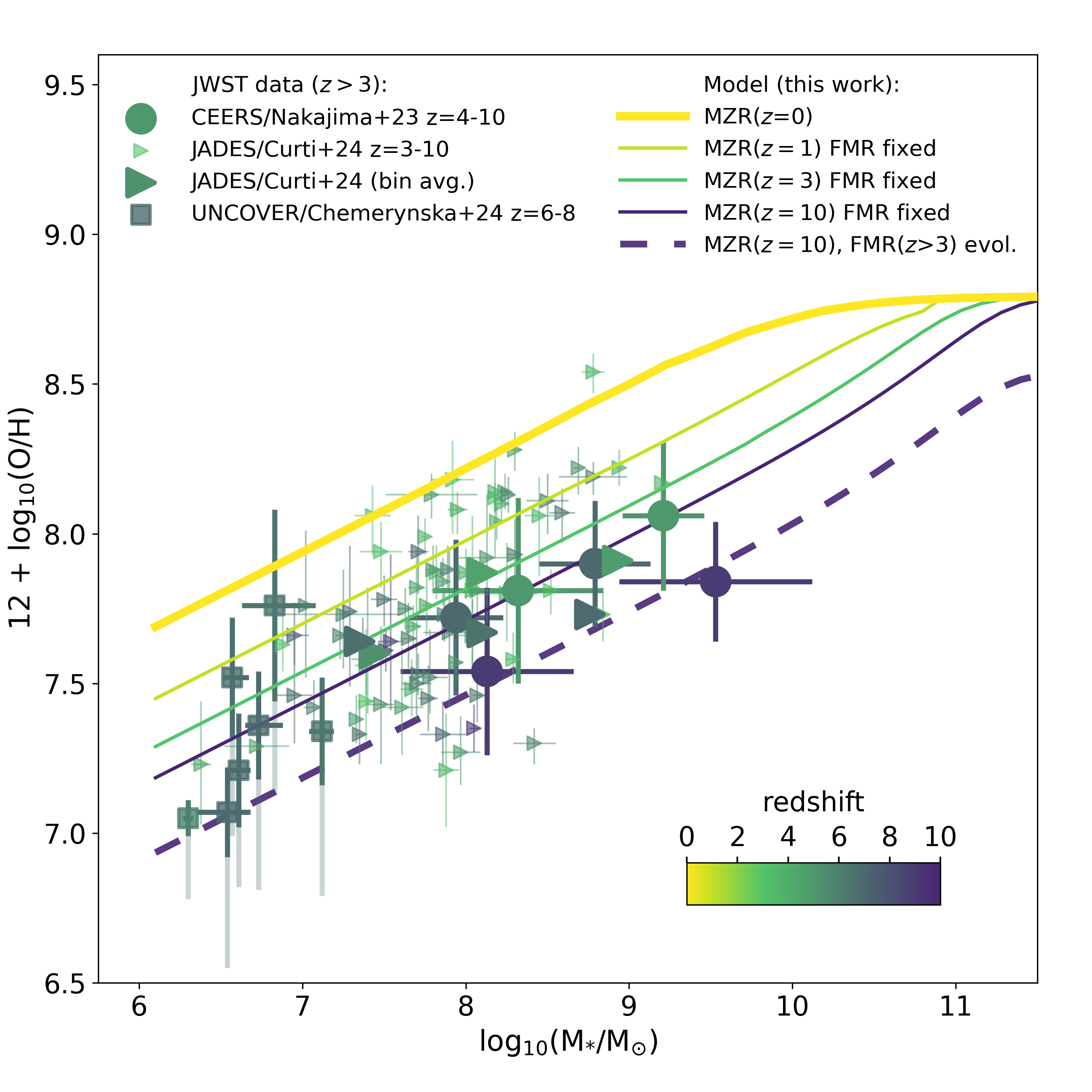}
\caption{ 
Evolution of the gas-phase $Z_{O/H}$-MZR in our model against $z>3$ data.
MZR($z$>3) predicted based on the known evolution of the SFMR and $z$-invariant FMR (modelled as in \citetalias{Chruslinska21}) is consistent with the current data, but additional evolution may be present. Solid lines - MZR at $z$=0, 1,3 10 assuming a $z$-invariant FMR. Dashed line - MZR at $z$=10, assuming additional evolution in the FMR normalisation at $z>3$. 
Data points - $z>3$ samples from \cite{Nakajima23} (circles, $z$-bin average), \citet{Curti24} (triangles, small - individual, big - $z$-bin average) and \citet{Chemerynska24} (squares, assuming calibration from \cite{Nakajima22}, faint errors extend to values estimated using \cite{Sanders24} calibration). Data points are coloured by redshift.
 Shown for a variation described in Sec. \ref{sec: method}, but for MZR($z$=0) from \cite{Curti20} (i.e. without +0.2 dex offset, to compare with the data as reported).
}
\label{fig: oxygen-MZR}
\end{figure}

The high $z > 3$ regime has only recently become accessible to empirical $Z_{O/H}$ studies, with the launch of the JWST enabling the detection of the emission lines conventionally used to determine the gas-phase metallicity.
Here we take advantage of these new constraints to update our framework and validate assumptions about the $Z_{O/H}$ evolution made in \citetalias{ChruslinskaNelemans19} (MZR with a strong $z$ dependence) and \citetalias{Chruslinska21} ($z$-invariant FMR) .
These two approaches give consistent f$_{\rm SFR}$(Z$_{O/H},z$) up to $z\lesssim$3, but lead to vastly different outcomes at higher $z$ (\citealt{Boco21},\citetalias{Chruslinska21}).\\
 We model the evolution of the MZR assuming that the FMR is $z$-invariant and compare it with the recent $z>3$ observational samples \citep{Nakajima23,Curti24,Chemerynska24} in Figure \ref{fig: oxygen-MZR}.
We follow the phenomenological description of the FMR developed in \citetalias{Chruslinska21}, where the FMR is determined by the local MZR($z=0$) and SFMR($z=0$) relations and a single parameter $\nabla_{FMR0}$. This approach captures the systematic differences in the FMR introduced by Z$_{O/H}$ and SFR determination methods \citep[e.g.][]{KewleyEllison08,Yates12,Kashino16,Telford16,MaiolinoMannucci19}.
MZR($z$) shown in Fig. \ref{fig: oxygen-MZR} as solid lines correspond to maximum density regions occupied by our model galaxies in the Z$_{O/H}$ - log$_{10}$(M$_{*}$) plane at fixed $z$. This locus changes with $z$ due to evolution of both SFMR and GSMF. We assume the MZR($z=0$) from \cite{Curti20} (yellow solid line), SMFR and GSMF as described in Sec. \ref{sec: method} \footnote{MZR evolution is similar for all variations considered here, but see \citetalias{Chruslinska21} and Fig. A4 therein for dependence on the MZR($z$=0) and $\nabla_{\rm FMR0}$.}. 
\\
The resulting MZR shows a strong evolution up to $z\approx$1.5 (in the last $\approx$9.5 Gyr), by which point the average Z$_{O/H}$ of log(M$_{*}$/M$_{\odot}$)$\lesssim$9 galaxies decrease by $\approx$0.3 dex. 
This is broadly consistent with observational estimates, showing that the average Z$_{O/H}$ of galaxies at $z\approx 1\text{--}3$ are lower than estimated from $z=0$ MZR by about 0.3 dex 
\citep[e.g.][]{Sanders20,Topping20,Cullen21,Henry21,Revalski24,Stanton24}.
Evolution within $z= 1\text{--}3$ range is difficult to quantify empirically as it is comparable to measurement uncertainties \citep[e.g.][]{Henry21} and smaller than systematics (e.g. \citealt{Laseter24,Stanton24}) introduced by different high-$z$ strong line metallicity calibrations used in the literature (e.g. \citealt{Bian18,Curti20,Nakajima23,Sanders24})
\\
At $z>3$, the MZR in our universal FMR model decreases further by $\approx$0.2 dex by $z\approx6$ and shows little evolution beyond this point. 
Observationally, the exact evolution of the MZR 
between $z = 3 \text{--} 10$ is currently unclear. 
\citet{Curti24} analyse a sample of $z=$3-10 galaxies from JADES \citep{Bunker23} and CEERS \citep{Finkelstein23} surveys, finding $\sim$0.05 dex -- 0.25 dex offset with respect to z$\sim$3 MZR (depending on the reference $z=3$ relation). This is consistent with \cite{Nakajima23}, who point out lower average Z$_{O/H}$ in their $z = 8 \text{--} 10$ bin, but consistent with no MZR evolution between $z = 4 \text{--} 10$ within errors.
Both \cite{Nakajima23} and \cite{Curti24} find weaker Z$_{O/H}$ evolution at the low $M_{*}$ end. However, this has been challenged by \cite{Chemerynska24} who analysed the low $M_{*}$ galaxy sample from the UNCOVER survey \citep{Bezanson24}.\\
 The picture emerging from current observations is that the normalisation of the MZR at $z = 3 \text{--} 10$ is somewhere $\approx$0.4-0.7 dex below that of the $z=0$ relation \citep{Curti24,Nakajima23,Morishita24}. 
 Even with this uncertainty, it is clear that the extrapolation used in \citetalias{ChruslinskaNelemans19}, leading to $>$1 dex decrease in Z$_{O/H}$ between $z=3$ and $z=10$, severely underestimates galaxy Z$_{O/H}$ at high $z$.
In turn, the $z$-invariant FMR from \citetalias{Chruslinska21} to MZR($z$) which is  broadly consistent with current observational estimates up to $z=10$.
However, stronger evolution is also allowed within current uncertainties (i.e., the universal FMR may lead to overestimated Z$_{O/H}$ at $z>6$).
To account for this possibility, we introduce a variation of our model in which the FMR normalisation decreases by 0.25 dex between $z=3$ and $z=10$. For simplicity, we assume that this decrease is linear with $z$ (see Tab. \ref{tab: variations}). The resulting MZR($z=10$) is shown in Fig. \ref{fig: oxygen-MZR} as a dashed line.

\section{Iron-dependent cosmic star formation history: example variation}\label{sec: results-example variation}

 \begin{figure*}[h]
\sidecaption
  \includegraphics[width=12cm]{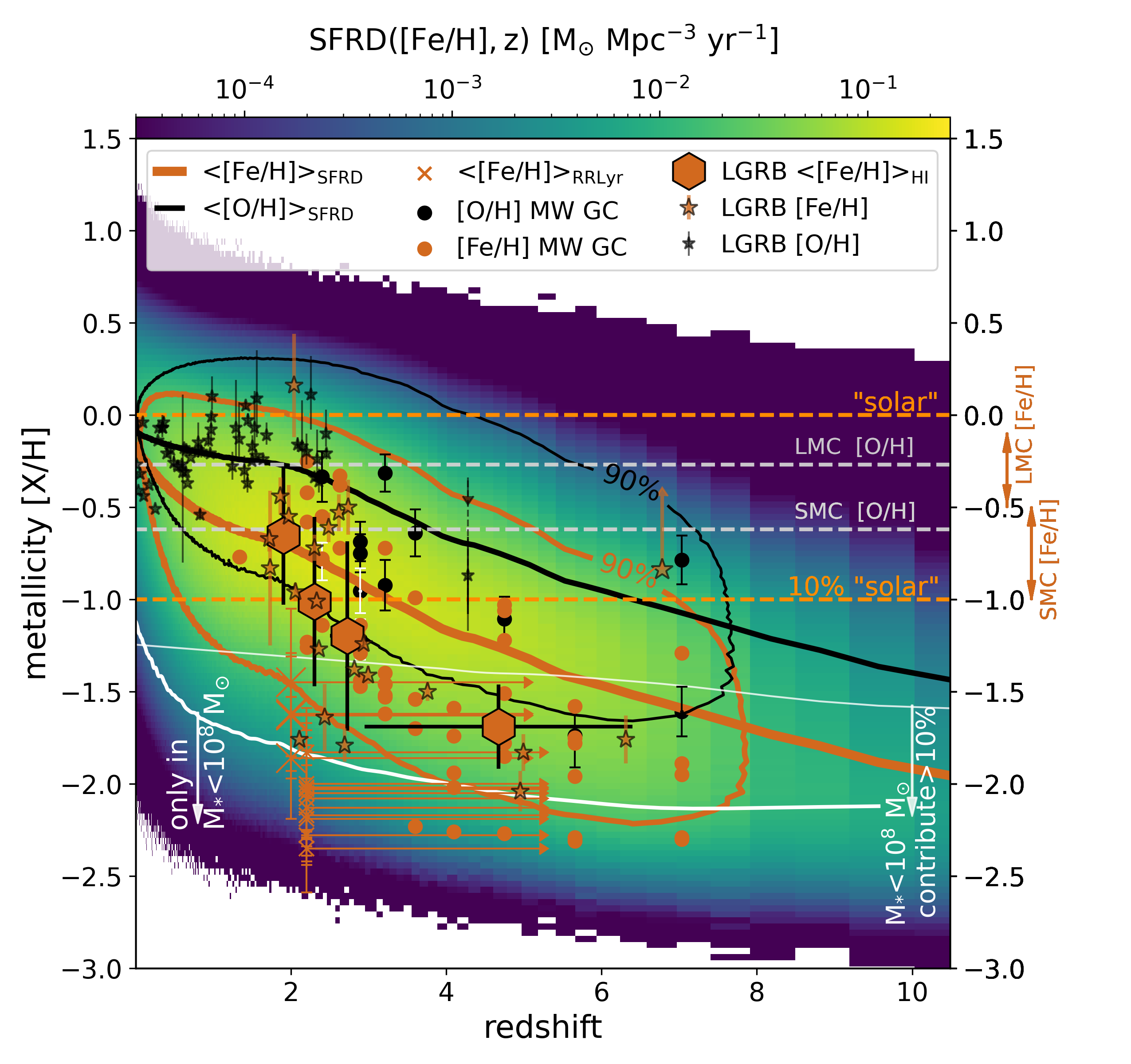}
     \caption{Distribution of the cosmic SFRD at different [Fe/H] and $z$ (colour scale), shown for an example variation described in Sec. \ref{sec: method}.
Black/brown lines: SFRD-weighted average [O/H] / [Fe/H] at each $z$. Black/brown contours enclose 90\% of the SFRD
to further show the offset between the $f_{\rm SFR}$([O/H],$z$) and $f_{\rm SFR}$([Fe/H],$z$).
White solid lines: [Fe/H] below  which  $M_{*}<10^{8}\Msun$ galaxies contribute at least 10\% of the SFRD (thin line) and below which contribution of $M_{*}>10^{8}\Msun$ galaxies is negligible (thick line). 
Data points: cross-check samples introduced in Sec. \ref{sec: Fe-cross-check-samples}.
Hexagons: $\langle \mathrm{[Fe/H]} \rangle_{\rm HI}$ of LGRB hosts binned in $z$. Brown/black stars: [Fe/H]/[O/H] of individual LGRB hosts.
Crosses: mean [Fe/H] of RR Lyrae stars in local dwarf galaxies (small/big: $M_{*}<10^{8}\Msun$ / $M_{*}\gtrsim 10^{8}\Msun$, offset for visibility) but formed at $z>2-5$ (horizontal arrows). 
Brown/black dots: [O/H]/[Fe/H] of MW globular clusters.
[O/H] and [Fe/H] of the LMC and SMC on our solar scale are indicated for the reference.}
    \label{fig: SFRD_Z_z_example}
\end{figure*}

We begin the discussion of our results by highlighting the key differences between iron- and oxygen-based metallicity-dependent cSFH.
Figure \ref{fig: SFRD_Z_z_example} shows the $f_{\rm SFR}(Z_{Fe/H},z)$  (plotted as a 2-dimensional histogram in the [Fe/H] - $z$ plane, coloured by the SFRD) for our example variation described in Section \ref{sec: method}. 
The corresponding $f_{\rm SFR}(Z_{O/H},z)$ is shown in Fig. \ref{fig: SFRD_OH_z_example}. 
To compare the iron- and oxygen-based results, we plot the SFRD-weighted averages of [Fe/H] and [O/H] at each $z$ (solid lines: $\langle \mathrm{[Fe/H]} \rangle_{\rm SFRD}$ - brown; $\langle \mathrm{[O/H]} \rangle_{\rm SFRD}$ - black).
We note that:
\begin{enumerate}[(i)]
\item The SFRD-weighted average metallicity is lower when iron is used as a metallicity probe instead of the oxygen, reflecting the delay in iron enrichment relative to oxygen. Relatedly, the distribution f$_{\rm SFR}(Z_{Fe/H},z)$ extends to lower metallicities than f$_{\rm SFR}(Z_{O/H},z)$.
In Sec. \ref{sec: solar O/Fe} we show that their offset is sensitive primarily to the [O/Fe] $-$ sSFR relation.
\item The offset between the $\langle \mathrm{[Fe/H]} \rangle_{\rm SFRD}$ and $\langle \mathrm{[O/H]} \rangle_{\rm SFRD}$ increases with $z$ until $z\approx3$ where $\langle \mathrm{[Fe/H]} \rangle_{\rm SFRD}$ shows a steeper evolution. At $z>3$ this offset approaches a constant value. This is a consequence of the weak evolution of galaxies' sSFRs at $z\gtrsim 3$, and thus, their locations in the [O/Fe] $-$ sSFR plane.
\item The mismatch between the two distributions suggests that the star formation with a solar O/Fe abundance ratio is rare throughout cosmic history (see Fig. \ref{fig: f_SFR_OFe_t_mixed}, showing the corresponding [O/Fe]-dependent cosmic SFH).
\end{enumerate}
We further quantify point (iii) in Section \ref{sec: solar O/Fe}. 

\subsection{Cross-check with observational samples} \label{sec: Fe-cross-check-samples}
While the f$_{\rm SFR}(Z_{O/H},z)$ is obtained by combining the empirical properties of galaxy populations across $z$, the f$_{\rm SFR}(Z_{Fe/H},z)$ is inferred more indirectly. The $Z_{\rm Fe/H}$ of young stars/gas is estimated from the galaxy's $Z_{\rm O/H}$ and sSFR via the [O/Fe]--sSFR relation.
To assess whether our results fall within a reasonable range, we compare them against three reference observational samples that probe different regions of $f_{\rm SFR}(Z_{\rm Fe/H}, z)$, as discussed below and shown in Fig. \ref{fig: SFRD_Z_z_example}.

\subsubsection{RR Lyrae:  [Fe/H] in low-mass, high-$z$ galaxies}

RR Lyrae are pulsating stars with typical ages of $\gtrsim 10$ Gyr, reaching $\gtrsim 12.5$ Gyr in GCs and in the lowest-mass dwarf galaxies \citep{Catelan04,Catelan09,Sesar14}. They trace the properties of the oldest stellar populations. Following \citet{Bellazzini25}, we adopt the mean metallicity of RR Lyrae stars in local dwarf galaxies ($\langle \mathrm{[Fe/H]} \rangle_{\rm RRLyr}$) as a proxy for the iron enrichment of the lowest-mass galaxies ($M_{*}<10^{8}\,M_\odot$) at $z \gtrsim 2$–5 (lookback times $\gtrsim$10 - 12.5 Gyr). Specifically, we use the values listed in Table A.1 of \citet{Bellazzini25}, based on [Fe/H] estimates from \citet{Muraveva25}.
These were derived from the pulsation periods and Fourier decomposition parameters of RR Lyrae light curves in Gaia Data Release 3 \citep{Clementini23}. 
\\
As a minimum requirement, any viable model must reproduce the formation of stars with $\langle \mathrm{[Fe/H]} \rangle_{\rm RRLyr}$ at $z \gtrsim 2$, in the regime where galaxies with $M_{*}<10^{8}\,M_\odot$ contribute to $f_{\rm SFR}(Z_{\rm Fe/H}, z)$.
Fig. \ref{fig: SFRD_Z_z_example} shows that this is the case, and that the full extent of the low [Fe/H]-high $z$ tail of the f$_{\rm SFR}(Z_{Fe/H},z)$ probed by RR Lyrae can only be populated by star formation in $M_{*}<10^{8} \Msun$ galaxies.
Conversely, not accounting for such galaxies leads to negligible star formation at [Fe/H]<-2 at any $z$ covered by our models.

\subsubsection{MW globular clusters: [Fe/H] at high SFRD}
Globular clusters (GCs) are old stellar systems, each thought to have formed in a brief star formation episode. While they host multiple stellar populations distinguished by variations in their light-element abundances, they generally exhibit only small spreads in $Z_{\rm Fe/H}$, likely reflecting the iron content of the interstellar medium at the time of their formation
\citep{McKenzie2021, Legnardi2022, Monty2023, Legnardi2024}. Both observations \citep[e.g.][]{Adamo20_local,Adamo20} and theory \citep[e.g.][]{grudic21} indicate that GC progenitor formation efficiency depends on the star-forming environment. Dense and most actively star-forming regions (such as found in local starbursts or in the early Universe) favour the formation of massive, long-lived clusters that can evolve into today’s GCs \citep{Adamo20,ReinaCampos22,Kruijssen25} 
Thus, we expect the GCs to preferentially probe the regions of $f_{\rm SFR}(Z_{\rm Fe/H}, z)$ with high-SFRD and (because of the FMR) low $Z_{\rm Fe/H}$, while avoiding the extremes of the distribution.
\\
In Fig. \ref{fig: SFRD_Z_z_example}, we plot the sample of MW GCs with isochronal ages and $Z_{\rm Fe/H}$ from \cite{Vandenberg13}. For each cluster, we assume its best-fit age corresponds to the lookback time/$z$ of its formation. We also show $Z_{\rm O/H}$ measurements for 16 MW GCs from \cite{Carretta09}, noting that GC stars can display large internal spreads in this element. We adopt the $Z_{\rm O/H}$ values from \cite{Monty25} (shifted to our solar scale), who restrict the analysis to O-normal populations and apply a +0.15 dex correction to place the estimates on a common abundance scale with other MW samples (namely the GALAH survey \citealt{Buder21}), using \cite{Amarsi19} as a baseline for their detailed treatment of NLTE effects.
\\
Fig. \ref{fig: SFRD_Z_z_example} shows that most of the MW GCs ($Z_{\rm Fe/H}$: brown dots; $Z_{\rm O/H}$: black dots) fall within the corresponding $f_{\rm SFR}(Z_{\rm Fe/H}, z)$ and $f_{\rm SFR}(Z_{\rm O/H}, z)$ contours, which enclose the regions where the bulk of stellar mass formed. As expected, none of the GCs are found in the extreme tail of the distribution (dark blue area). 
However, a few GCs 
\footnote{NGC4590, NGC5053, NGC5466, NGC6341, NGC7078, NGC7099, and Terzan 8}
fall in the region where, according to our model, star formation occurs almost exclusively in galaxies with $M_{*}<10^{8}\,\Msun$. 
This is somewhat surprising, as such low-mass systems are generally not thought capable of (efficiently or at all) forming the massive (initially $\gtrsim10^{6}\,\Msun$) star clusters that are the progenitors of present-day GCs \citep{Renaud18,Krumholz19,Adamo20}. Interestingly, \emph{all} of the GCs which fall in this region have been tagged as accreted clusters based on their orbits, the chemistry of their O-normal population, or using a combination of both orbits and chemistry \citep{Massari2019, Massari2025, Belokurov2024, Monty2024}.
A more detailed discussion is beyond the scope of this paper, but we return to this comparison in our follow-up study.

\subsubsection{Long gamma ray burst hosts: (lower limit on the) $\langle \mathrm{[Fe/H]} \rangle_{\rm SFRD}$($z$)} \label{sec: LGRB sample}

Long-duration GRBs are typically linked to the deaths of massive stars and are found in actively star-forming regions \citep[e.g.][]{Galama_1998,Hjorth03, Christensen_2004, Fruchter_2006}. They are among the most energetic electromagnetic transients known and have been detected out to $z \gtrsim 8$ \citep{Tanvir09}, probing star-forming environments across cosmic time. 
LGRB afterglows allow determination of the gas-phase $Z_{\rm Fe/H}$ in their hosts via absorption-line spectroscopy \citep[e.g.][]{Vreeswijk04}. This is especially valuable for our study, since direct constraints on $Z_{\rm Fe/H}$ in high-$z$ star-forming environments are scarce. Furthermore, unlike galaxies selected in conventional surveys, LGRB-afterglow selected systems are not expected to be biased against faint objects.
\\
We use absorption-derived LGRB host properties (total metallicity - $[M/H]_{\rm tot}$, $z$, and HI column density - $N_{\rm HI}$) compiled by \citet{Heintz23}, covering $1.7 < z < 6.3$. We also include GRB 050505 at $z \approx 4.28$ (\citealt{Berger06}, as reanalysed by \citealt{Inkenhaag25}), and GRB 240218A at $z \approx 6.78$, for which, however, only the lower limit of $[M/H]_{\rm tot}\geq -0.8$ is available \citep[][revised analysis]{Saccardi25}.
Importantly, $[M/H]_{\rm tot}$ represents the total metallicity (gas + dust) of the ISM, corrected for dust depletion \citep[e.g.][]{DeCia16,DeCia18}.
This correction is critical for iron, which is strongly depleted into dust grains (by 90-99\% in the MW e.g. \citealt{Jenkins09}, and up to 90\% in lower-metallicity environments, e.g. \citealt{DeCia16,Konstantopoulou24}).
$[M/H]_{\rm tot}$ is equivalent to a dust-corrected [Fe/H] when Zn abundance is available to guide the dust correction. 
If this is not the case, S or Si can be used for dust corrections, but such estimates are more approximate and the resulting $[M/H]_{\rm tot}$ may overestimate $Z_{\rm Fe/H}$ unless $\alpha$-enhancement is explicitly accounted for \citep{Saccardi23,DeCia24}.
We assume that the $[M/H]_{\rm tot}$ values obtained with Zn-based dust corrections (i.e. GRB 240218A, and 21 of the 36 objects from \citealt{Heintz23}) provide a reliable proxy for $Z_{\rm Fe/H}$. We also assume that this is the case for GRB 210905A, as its $[M/H]_{\rm tot}$ accounts for $\alpha$-enhancement \citep{Saccardi23}.\\
We plot the [Fe/H] estimates for individual LGRB hosts in this sample in  Fig. \ref{fig: SFRD_Z_z_example} and we use those estimates (without the lower limit from GRB 240218A) to re-derive the $N_{\rm HI}$-weighted mean $\langle \mathrm{[Fe/H]} \rangle_{\rm HI}$ with $z$, following Eq. 2 of \citet{Heintz23} (see Tab. \ref{tab:LGRB binned} in the appendix for the binning and values).
In Sec. \ref{sec: Fe/H avg. comparison} we also show $\langle \mathrm{[M/H]}\rangle_{\rm HI}$ obtained for the full sample.
Weighting by $N_{\rm HI}$ accounts for the varying amounts of potentially star-forming gas in the probed galaxies.
This enables a comparison between $\langle \mathrm{[Fe/H]} \rangle_{\rm HI}(z)$ and the $\langle \mathrm{[Fe/H]} \rangle_{\rm SFRD}(z)$ derived from our models.
 However, the comparison is complicated by the fact LGRB progenitors are theorized to be massive, iron-poor stars (\citealt{Hirschi_2005,YoonLanger05,WoosleyHeger06}, but see \citealt{Chrimes20,Briel25}). Thus, they may preferentially occur in regions of high star formation and low $Z_{\rm Fe/H}$. Empirical studies indicate a decline in LGRB occurrence near solar metallicity, but overall find no clear cutoff or preference for extremely metal-poor environments \citep[e.g.][]{Vergani15,Japelj16,Perley16, Graham23,Disberg25} and at intermediate redshifts, LGRB host galaxies appear to trace the star-forming population \citep[][selected by having cold gas and active starformation]{Krogager24}.
In the presence of such biases, the LGRB-based $\langle \mathrm{[Fe/H]} \rangle_{\rm HI}$ provides a lower limit on the overall $\langle \mathrm{[Fe/H]} \rangle_{\rm SFRD}$ and we expect the offset between the two to decrease toward high $z$, where galaxies have higher SFRs and lower $Z_{\rm Fe/H}$.
\\
Finally, while we focus on $Z_{\rm Fe/H}$, in Fig. \ref{fig: SFRD_Z_z_example} we also show  $z \lesssim 2.5$ LGRB hosts with emission-line $Z_{\rm O/H}$ estimates from \cite{Graham23} (obtained with \citealt{Curti17,Curti20} calibration, shifted by +0.2 dex for consistency with our scale) and GRB 050505 \citep[][]{Inkenhaag25}, for which we show the range of $Z_{\rm O/H}$ obtained with different methods. 
We note that samples with emission-line based $Z_{\rm O/H}$ are subject to the usual selection effects, favouring brighter (more massive, lower redshift) sources. We refer the interested reader to \citet{Disberg25} for a recent detailed discussion of the comparison with this sample.
\\
Based on the comparison shown in Fig. \ref{fig: SFRD_Z_z_example} we note the following:
\begin{itemize}
\item $Z_{\rm Fe/H}$ and $Z_{\rm O/H}$  of individual LGRB hosts show appreciable scatter, which our model fully accounts for.
\item  $\langle \mathrm{[Fe/H]} \rangle_{\rm HI}$ from absorption-derived LGRB host properties does not trace $\langle \mathrm{[O/H]} \rangle_{\rm SFRD}$. However, $\langle \mathrm{[Fe/H]} \rangle_{\rm SFRD}$ is broadly consistent with $\langle \mathrm{[Fe/H]} \rangle_{\rm HI}$.
\end{itemize}
The second point demonstrates the importance of accounting for non-solar abundance ratios when comparing "metallicities" probed by different elements, particularly in the context of the potential low metallicity bias of LGRB. Our model predicts a milder evolution than the LGRB-based estimate in the highest-$z$ bin; however, $\langle \mathrm{[Fe/H]} \rangle_{\rm HI}$ estimate in this bin is likely unreliable (see Sec.~\ref{sec: Fe/H avg. comparison}, where we also compare across model variations), and we refrain from drawing conclusions from comparison in this $z$ range.

\section{How (un)common is the solar abundance pattern?} \label{sec: solar O/Fe}
\begin{figure}[htbp]
    \centering
    \includegraphics[width=\columnwidth]{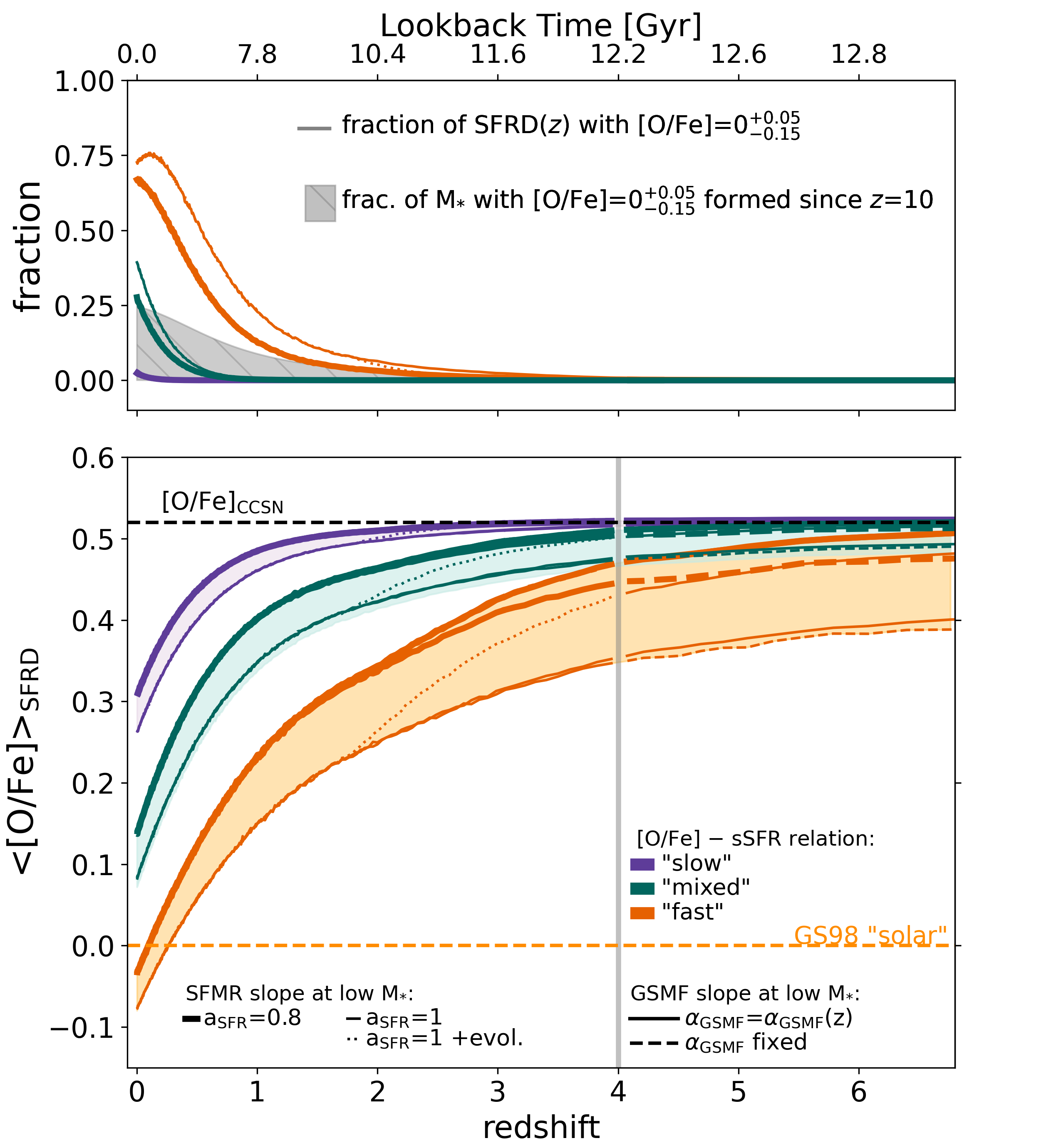}    
    \caption{
Top: The gray area indicates that, across our models, the fraction of cosmic stellar mass formed since $z$=10 with near-solar [O/Fe] (between –0.15 and 0.05 dex) is at most 25\%. Lines show the fraction of the total SFRD($z$) formed with near-solar [O/Fe].
Bottom: SFRD weighted mean [O/Fe] as a function of redshift/lookback time. 
At $z<4$ ($z>4$), the line styles distinguish between the  assumptions about the SFMR ($\alpha_{\rm GSMF}$). colours indicate the [O/Fe]--sSFR relation and the coloured areas indicate the range spanned by models with a fixed [O/Fe]--sSFR relation.
Green dotted line corresponds to our example variation.
}
    \label{fig: OFe_avg}
\end{figure}

The example discussed in Sec. \ref{sec: results-example variation} shows a persistent offset between $\langle \mathrm{[O/H]} \rangle_{\rm SFRD}$ and $\langle \mathrm{[Fe/H]} \rangle_{\rm SFRD}$, i.e. the SFRD-weighted mean $\langle \mathrm{[O/Fe]} \rangle_{\rm SFRD} > 0$ at all redshifts.
This indicates that most of the cosmic star formation happened with O/Fe ratio larger than `solar'.
We quantify this result and its sensitivity to assumptions about the uncertain factors in our model in Fig. \ref{fig: OFe_avg}.
\\
The [O/Fe]–sSFR relation represents the dominant source of uncertainty in the  $\langle \mathrm{[O/Fe]} \rangle_{\rm SFRD}$ and in the fraction of the SFRD with `near-solar' O/Fe abundance ratios, liberally defined here as a range between [O/Fe]=-0.15 and 0.05 dex (log$_{10}$(O/Fe)=1.18 - 1.38) to enclose different reference solar abundance ratios used in the literature \footnote{$Z_{\rm Fe/H}$ = 7.45 - 7.5 and $Z_{\rm O/H}$=8.69 - 8.83 \citep{Lodders09,Asplund09,Asplund21,GrevesseSauval98}, leading to -0.15 to +0.05 offsets in [O/Fe] relative to our scale.}.
colours in Fig.  \ref{fig: OFe_avg} correspond to different average [O/Fe]–sSFR relations, varied between the extreme cases discussed in Sec. \ref{sec: which O/Fe-sSFR}.
The $\langle \mathrm{[O/Fe]} \rangle_{\rm SFRD}$ approaches the solar value only in the `fast' Fe-enrichment scenario (orange curves), which assumes the minimum possible delay between CCSN and the onset of enhanced Fe enrichment. Even in this case, $\langle \mathrm{[O/Fe]} \rangle_{\rm SFRD}\lesssim$0 only at $z \lesssim 0.4$.
On the other extreme, using the  [O/Fe]–sSFR relation corresponding to the `slow' Fe-enrichment scenario (purple curves) leads to $\langle \mathrm{[O/Fe]} \rangle_{\rm SFRD}\gtrsim$0.25 dex even at $z$=0 and the fraction of SFRD($z=0$) with `near-solar' O/Fe ratio is $\lesssim$2\%.
We note that the fraction of cosmic stellar mass formed with near-solar [O/Fe] is the most sensitive to the [O/Fe]–sSFR relation around $\log_{10}({\rm sSFR})\sim -9$ (i.e., main-sequence galaxies at $z\sim0.6$–2). Only a steeper drop to solar-like [O/Fe] in this regime can visibly increase this fraction. However, a relation steeper in this regime than our extreme “fast” Fe-enrichment scenario would challenge the current constraints on the relation \citepalias[see Fig. \ref{fig: OFe-sSFR rel.} and][]{Chruslinska24_OFe}. Variations in the shape of the relation at lower sSFR have only a marginal impact, because such galaxies contribute little to the (integrated) cosmic SFRD budget.
\\
Another factor important for those results is the distribution of galaxy sSFR: a higher fraction of galaxies with high sSFR increases $\langle \mathrm{[O/Fe]} \rangle_{\rm SFRD}$, whereas weaker evolution in galaxy sSFR with $z$ leads to weaker evolution of  $\langle \mathrm{[O/Fe]} \rangle_{\rm SFRD}$.
To quantify this effect, in Fig. \ref{fig: OFe_avg} we consider three variations of the SFMR (shown in Fig. \ref{fig: scaling rel.}): the relation from \citetalias{Popesso23} (thin solid lines), the same relation but with a shallower slope $a_{\rm SFR}$=0.8 instead of $a_{\rm SFR}$=1 at low $M_{*}$ (thick solid lines), and a version with increased high-$z$ evolution (dotted lines; as in the example in Sec. \ref{sec: results-example variation}). We vary $a_{\rm SFR}$ because (i) our analysis extends to galaxies with  M$_{*}$ much lower than log$_{10}$(M$_{*}$/M$_{\odot}$)>8.7 adopted in \citetalias{Popesso23} to restrict their fit to regimes with high sample completeness, and (ii) $a_{\rm SFR}$ has a strong impact on our results: adopting $a_{\rm SFR}$=0.8 instead of $a_{\rm SFR}$=1 leads to $\sim$1 dex higher log$_{10}$(SFR) at $M_{*}=10^{6} \Msun$.
 While observational studies report even lower $a_{\rm SFR}$ values \citep[particularly at high $z$, see e.g. compilations in][]{Speagle14,Boogaard18,Merida23}, \citet{Simmonds25} attribute these to sample incompleteness, while \citet{Leja15}  argue that such shallow slopes are inconsistent with the observed evolution of the GSMF (see also \citealt{DiCesare25}). Moreover, assuming $a_{\rm SFR}$<0.8 would cause a significant overestimate of the cSFH (see Sec. \ref{sec: cSFH}).
 The effect of $a_{\rm SFR}$ and/or galaxy sSFR evolution on  $\langle \mathrm{[O/Fe]} \rangle_{\rm SFRD}$ can be partially offset by the abundance of low-$M_{*}$ galaxies. In our framework, the latter is tied to uncertain low-$M_{*}$ slope of the GSMF. Assuming fixed $\alpha_{\rm GSMF}$ (lines continued as dashed at $z>4$) versus $\alpha_{\rm GSMF}(z)$ that steepens with $z$ has a relatively small effect on the results shown in Fig. \ref{fig: OFe_avg}. 
 Finally, we point out that:
 \begin{itemize}
     \item  The fraction of mass formed in stars until the present day with [O/Fe] between $-$0.15 dex and 0.05 dex is at most 25\% (gray band, top panel in Fig. \ref{fig: OFe_avg}).
     \item The fraction of cosmic SFRD happening with 'near-solar' [O/Fe] is negligible at $z>2$ across our models. This fraction is negligible even at $z\sim$0 if the delay between the CCSN enrichment and enhanced Fe production is long (O(10$^{8}$yr)).
 \end{itemize}
The exact values shown in the top panel of Fig.~\ref{fig: OFe_avg} depend to some extent on the assumed metallicity scatter within galaxies. We discuss this further in Appendix~\ref{app: scatter}, where we derive a more conservative upper bound of $\lesssim 30\%$ for the fraction of stellar mass formed with near-solar $[\mathrm{O/Fe}]$, adopting the extreme assumption that the stellar $Z_{\rm Fe/H}$ and $Z_{\rm O/H}$ distributions within a host are fully independent. We therefore conclude that the specific treatment of scatter does not affect our main result that star formation with near-solar $[\mathrm{O/Fe}]$ is rare.
This is in stark contrast to the fact that most stellar evolution and spectral synthesis models used in the literature, and applied to interpret observations, including those of young and high-$z$ environments, assume a solar abundance pattern (in particular, [O/H] $\approx$ [Fe/H] $\approx \log_{10}(Z/Z_{\odot})$).
We discuss the implications of this result in Sec. \ref{sec: implications}.

\subsection{The relevance of the solar abundance pattern} 
In this section, we argue that the results presented above are broadly consistent with (i) the Milky Way being representative of the star-forming galaxy population at solar formation epoch ( $\sim$4-5 Gyr ago or $z\approx0.4$), i.e. that a galaxy with Milky Way–like properties is a typical draw from the distributions introduced in Sec.~\ref{sec: method}, and (ii) the Sun’s abundances being common among stars formed in such systems at that time.\\
\begin{figure}[htbp]
\centering
\includegraphics[width=\columnwidth]{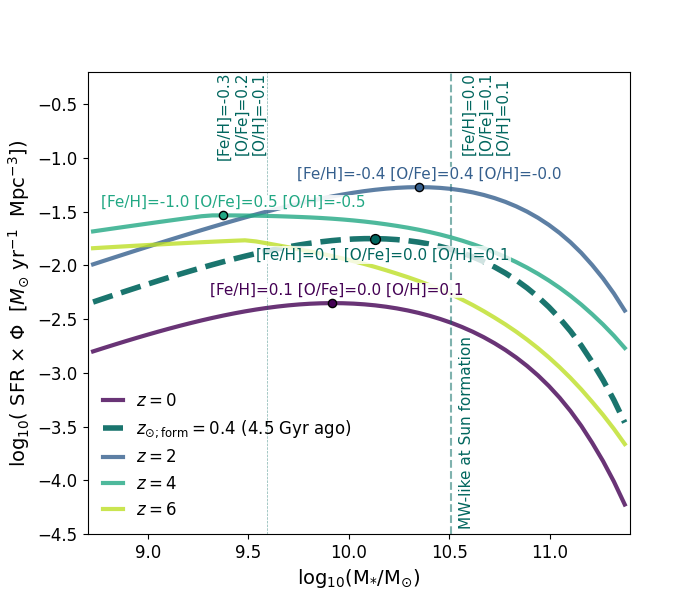}
\caption{
Contribution to the cosmic SFRD (SFR density per log $M_{*}$ bin) versus galaxy log$_{10}$(M$_{*}$) at several redshifts (colour-coded; bottom and top curves correspond to z=0 and z=2, respectively, illustrating the evolution of the cosmic star formation history).
Filled circles mark the log$_{10}$(M$_{*}$) at which the contribution peaks at each $z$ (ill-defined for $z>$4).
At each marked mass we list the typical abundances ([O/Fe],[O/H],[Fe/H]) of star-forming gas in our model.
The thick dashed curve highlights $z\approx$0.4, corresponding to a lookback time of 4.5 Gyr (the epoch of Solar birth). Vertical dashed lines show log$_{10}$(M$_{*}$)
within $\pm$0.1 dex of the $z=0.4$ peak; abundances for these two masses are reported at the top. 
This mass range encompasses main-sequence galaxies with stellar masses comparable to that of the Milky Way at the epoch of solar formation (i.e. not its present-day mass), which were forming stars with near-solar abundance ratios.
Shown for our example variation.
}
\label{fig: typical galaxy}
\end{figure}
First, we verify that in our model, Milky Way–like galaxies (i.e. galaxies with masses of $\log_{10}(M_{*}/\Msun)\approx 10.4 - 10.5$ at $z=0.4$ and $\log_{10}(M_{*}/\Msun)\approx 9.9 - 10$ at $z=2$, e.g. \citealt{Patel13,Grand17}) are among the dominant contributors to the cosmic SFRD at $z \lesssim 2$, consistent with the picture that the Milky Way and its progenitors are representative $L_{\ast}$ systems. 
This contribution is computed by weighting the SFR of main-sequence galaxies of different masses by their number density $n_{\rm gal}$ (given by the GSMF). The result is shown in Fig.~\ref{fig: typical galaxy} at several redshifts, with the thick dashed curve indicating the epoch of solar formation ($z=0.4$).
We find a relatively broad range in $\log_{10}(M_{*})$ contributing comparably to the cosmic SFRD, with main-sequence galaxies with masses comparable to that of the Milky Way at the epoch of solar formation lying within $\pm 0.1$ dex of the peak (between the vertical dashed lines in Fig.~\ref{fig: typical galaxy}).
\\
Second, we examine the gas-phase abundances assigned to such galaxies at a lookback time of 4–5 Gyr in our model.
These are reported in the top-right corner of Fig.~\ref{fig: typical galaxy} for our example variation. 
Using the FMR, we find $[\mathrm{O/H}]\approx 0.1$ dex. Taken at face value, this would make the Sun ($[\mathrm{O/H}]=0$) appear mildly O-poor relative to the average abundances of stars formed at that epoch in Milky Way–like galaxies (adopting a different $Z_{O/H,\odot}$ than our reference value from \citet{GrevesseSauval98} would generally increase this offset).
However, a $\sim0.1$ dex offset in $Z_{O/H}$ is comparable to the intrinsic scatter of the MZR ($\sigma_{\rm MZR}\approx 0.1$ dex at high $M_{*}$), implying that Milky Way–like galaxies with $[\mathrm{O/H}]=0$ lie within the typical population at $z=0.4$. Moreover, given the Sun’s Galactocentric distance of $\gtrsim 1.5,R_{e}$ and the negative $Z_{\rm O/H}$ gradients commonly observed in spirals, it may have formed at a lower $Z_{\rm O/H}$ than the global value inferred from galaxy scaling relations (assumed to be representative of the average at $R_{e}$). \citetalias{Chruslinska24_OFe} estimate a possible $\sim -0.05$ dex offset due to the Milky Way gradient \citep{Arellano-Cordova20}, comparable to the abundance spread observed among solar-like (i.e. similar in age and [Fe/H]) disk stars \citep{Nissen15}.
\\
The corresponding $[\mathrm{O/Fe}]$ for a main-sequence galaxy with Milky Way–like mass at $z=0.4$ follows from the $[\mathrm{O/Fe}]$–sSFR relation, and $[\mathrm{Fe/H}]=[\mathrm{O/H}]-[\mathrm{O/Fe}]$. In our “mixed” Fe-enrichment scenario (assumed in Fig.~\ref{fig: typical galaxy}), this yields $[\mathrm{O/Fe}]=0.1$ dex and $[\mathrm{Fe/H}]=0$. Adopting alternative relations leaves $[\mathrm{O/H}]$ unchanged but shifts both $[\mathrm{Fe/H}]$ and $[\mathrm{O/Fe}]$ by $\pm 0.2$ dex, making the Sun appear either Fe-poor (“fast” enrichment) or Fe-rich (“slow” enrichment) relative to the average abundances of its host galaxy at the solar formation epoch.
The intrinsic scatter in the $Z_{\rm Fe/H}$–MZR for Milky Way–mass galaxies, as well as the Milky Way’s radial gradient, are comparable to those for $Z_{\rm O/H}$ \citep[e.g.][]{Galazzi05,Spina21}, implying an expected $\sim 0.1$ dex spread in average $Z_{\rm Fe/H}$ across the galaxy population. Within galaxies, the dispersion of the $Z_{\rm Fe/H}$ metallicity distribution function for solar-neighbourhood thin-disk stars at fixed age is $\sim 0.15$–0.2 dex \citep{Casagrande11,Hayden15,Buder21,Feuillet19}. Thus, even if the mean $[\mathrm{Fe/H}]$ of Milky Way–like galaxies at $z=0.4$ differs by $\pm 0.2$ dex, stars with solar $[\mathrm{Fe/H}]$ are still expected to form given the observed internal scatter.
\\
In principle, all our model variations predict the formation of stars with solar abundance ratios in MW-like galaxies at the solar formation epoch, within the expected scatter. 
However, such stars need not align with the peak of the metallicity distribution function in MW-like galaxies. The match is closest for [O/Fe]–sSFR relation intermediate between our "mixed" and "fast" Fe-enrichment scenarios, and poorest if the relation resembles the "slow" Fe-enrichment case. In the latter case, negative radial metallicity gradients would further exacerbate deviations from solar $Z_{\rm Fe/H}$ and $Z_{\rm O/H}$. Zooming out to cosmic volume, depending on the [O/Fe]–sSFR relation, stars with [O/Fe]=0 at the solar formation epoch would either coincide with the peak of the [O/Fe] distribution ("fast", Fig.  \ref{fig: f_SFR_OFe_t_fast} ), lie within the region enclosing 95\% of the cosmic SFRD ("mixed", Fig.  \ref{fig: f_SFR_OFe_t_mixed}), or fall outside the contours enclosing 99\% of the cosmic SFRD ("slow").
A detailed comparison of our work with the MW requires careful accounting for selection effects, completeness, and systematics in MW stellar surveys, as well as a more careful treatment of the metallicity spread inside galaxies in our model, which is beyond the scope of this paper. 
For the purposes of this discussion, we note that if the MW is representative of star-forming galaxies, and if the near-solar $Z_{\rm O/H}$ and $Z_{\rm Fe/H}$ values are typical of stars formed in such systems at $z\lesssim0.4$, then our 'slow' Fe-enrichment scenario is disfavoured.

\section{Cosmic star formation history: model variations} 
\subsection{Total cosmic star formation history}\label{sec: cSFH}

 \begin{figure*}[h!]
\sidecaption
  \includegraphics[width=12cm]{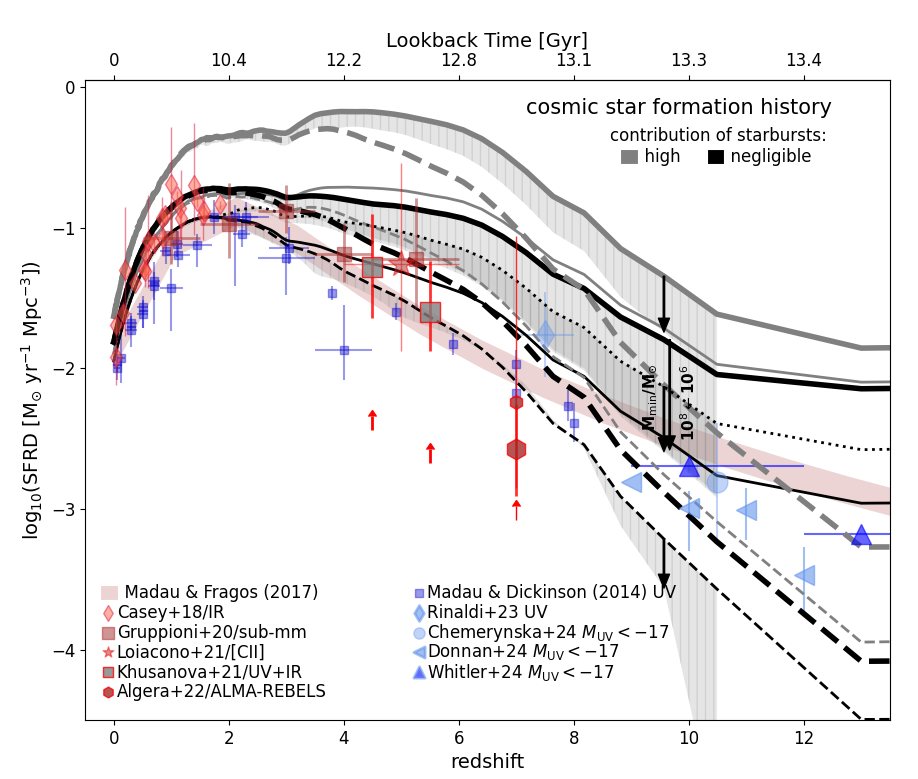}
     \caption{Cosmic star formation history for different combinations of assumptions about the GSMF (solid/dashed lines: $\alpha_{\rm GSMF}$ steepening with $z$/ fixed; see top panel in Fig. \ref{fig: scaling rel.}) and SFMR (thin/thick lines: \citetalias{Popesso23}  with $a_{\rm SFR}$=1/$a_{\rm SFR}$=0.8, dotted line: \citetalias{Popesso23} with additional evolution at $z>2$; see middle panel in Fig. \ref{fig: scaling rel.}). We also show the effect of different contribution of starbursts (black/gray lines - negligible/high). Data points - observational estimates (see legend), red band - estimate from \cite{MadauFragos17}.  Where necessary, we adjust the literature estimates to \cite{Kroupa01} IMF assumed here.
Hatched regions illustrate the effect of increasing the lower limit of integration over galaxy stellar mass from $10^{6}\Msun$ to $10^{8}\Msun$. Dotted line shows the cSFH of the variation considered in Sec. \ref{sec: results-example variation}.}
\label{fig: cSFH}
\end{figure*}

Figure \ref{fig: cSFH} shows the cSFH obtained under different combinations of assumptions in our model together with literature observational estimates. This serves as an important consistency check: although our framework is based on empirical relations, it does not include any normalisation to match independent constraints on the cSFH when these relations are combined.
The observational data shown include UV-based estimates
 (blue data points) compiled by \cite{MadauDickinson14}, and the recent $z>6$ estimates from \cite{Rinaldi22} (lower limit based on the galaxies in the COSMOS/SMUVS catalogue), \cite{Donnan24}, \cite{Whitler25} and \cite{Chemerynska24} (all three based on integration of the UV luminosity functions to the same limiting magnitude $M_{UV}=-17$).
Additionally, we show the IR/(sub)millimeter based estimates (red data points/limits) compiled by \cite{Casey18}, and the estimates from \cite{Gruppioni20}, \cite{Loiacono21}, \cite{Khusanova20} and \cite{Algera22}. These studies sample the contribution of dusty star-forming galaxies, constraining the obscured fraction of the SFRD to $z\sim$7. \citet{Martis25} probe the obscured fraction with photometry up to $z=9$, finding that up to $\sim$50\% of all SFR may still be obscured by at those early epochs, and thereby would be missed by the UV luminosity function-based SFRD estimates.
\\
As in the previous section, we vary the contribution of low-$M_{*}$ galaxies by changing $\alpha_{\rm GSMF}$ (top panel in Fig. \ref{fig: scaling rel.}) and $a_{\rm SFR}$ and we consider increased $z\gtrsim$2 evolution of the SFMR with respect to \citetalias{Popesso23} (middle panel in Fig. \ref{fig: scaling rel.}).
In gray we also show the cSFH obtained assuming a high SB contribution (see appendix \ref{sec, app: SB}), motivated by the results of \citet{Caputi17,Rinaldi22,Rinaldi25} who find a SB fraction f$_{SB}$(M$_{*}$, $z$) that strongly increases with $z$ and to low M$_{*}$.
\\
Direct comparison of our results with estimates from the literature is complicated by methodological differences. In particular, a consistent treatment would require mapping between the galaxy properties used in our model (SFR, M$_{*}$, metallicity) and observed galaxy luminosities, as well as homogenising assumptions regarding integration limits and extrapolations beyond sample completeness. Each of these aspects can introduce offsets of a factors of a few in the inferred SFRD, and their proper assessment lies beyond the scope of this work. We therefore adopt a conservative approach, not excluding any model variations that agree with literature SFRD estimates within a factor of 10.
These estimates are particularly sensitive to the assumed lower limit of the integration, whether in $M_{*}$ (our estimates) or in IR/UV luminosity (literature), where galaxy properties are uncertain.
To illustrate this, we show in Fig. \ref{fig: cSFH} (hatched regions) the effect of raising the lower integration limit from $M_{min}=10^{6}\Msun$ to $M_{min}=10^{8}\Msun$ for several variations. 
The impact grows with $z$, as does the apparent mismatch with literature estimates, reaching a factor of 3–5 at $z=8$ for models with evolving $\alpha_{\rm GSMF}(z)$ and $a_{\rm SFR}=0.8$ or with high SB contribution (both elevate SFR at $M_{*}<10^{9} \Msun$).
By contrast, for $\alpha_{\rm GSMF}$=$\alpha_{\rm fix}$, the effect is negligible to $z\gtrsim9$.\\
We note that a high SB contribution would lead to an overestimate of the SFRD at $z>2$ by a factor >10 compared to literature values, even for  $M_{min}=10^{8}\Msun$, unless a$_{\rm SFR}\gtrsim$1 (which is not the case in \citealt{Rinaldi22,Rinaldi25}). Such variations are not considered further.
The limiting case with high $f_{SB}$ and $a_{\rm SFR}=1$ (hereafter `high $f_{SB}$’) is, for our results, effectively indistinguishable from a scenario without an SB sequence but with $a_{\rm SFR}$ evolving with $z$ ($a_{\rm SFR}\sim0.9$ at $z=0$ and $a_{\rm SFR}\lesssim0.8$ at $z>2$). The corresponding results are qualitatively similar to those for the $a_{\rm SFR}=0.8$ case and, for clarity, we do not show them in all figures.
Finally, while our analysis does not extend beyond $z>$10.5, in Fig. \ref{fig: cSFH} we show the cSFH up to $z\sim 13$ to highlight its strong dependence on the extrapolation of the GSMF (see Appendix \ref{sec, app: GSMF}).

\subsection{"Low-metallicity" cosmic star formation history}\label{sec: low-Z cSFH}

\begin{figure*}[htbp]
    \centering
    \includegraphics[width=0.491\textwidth]{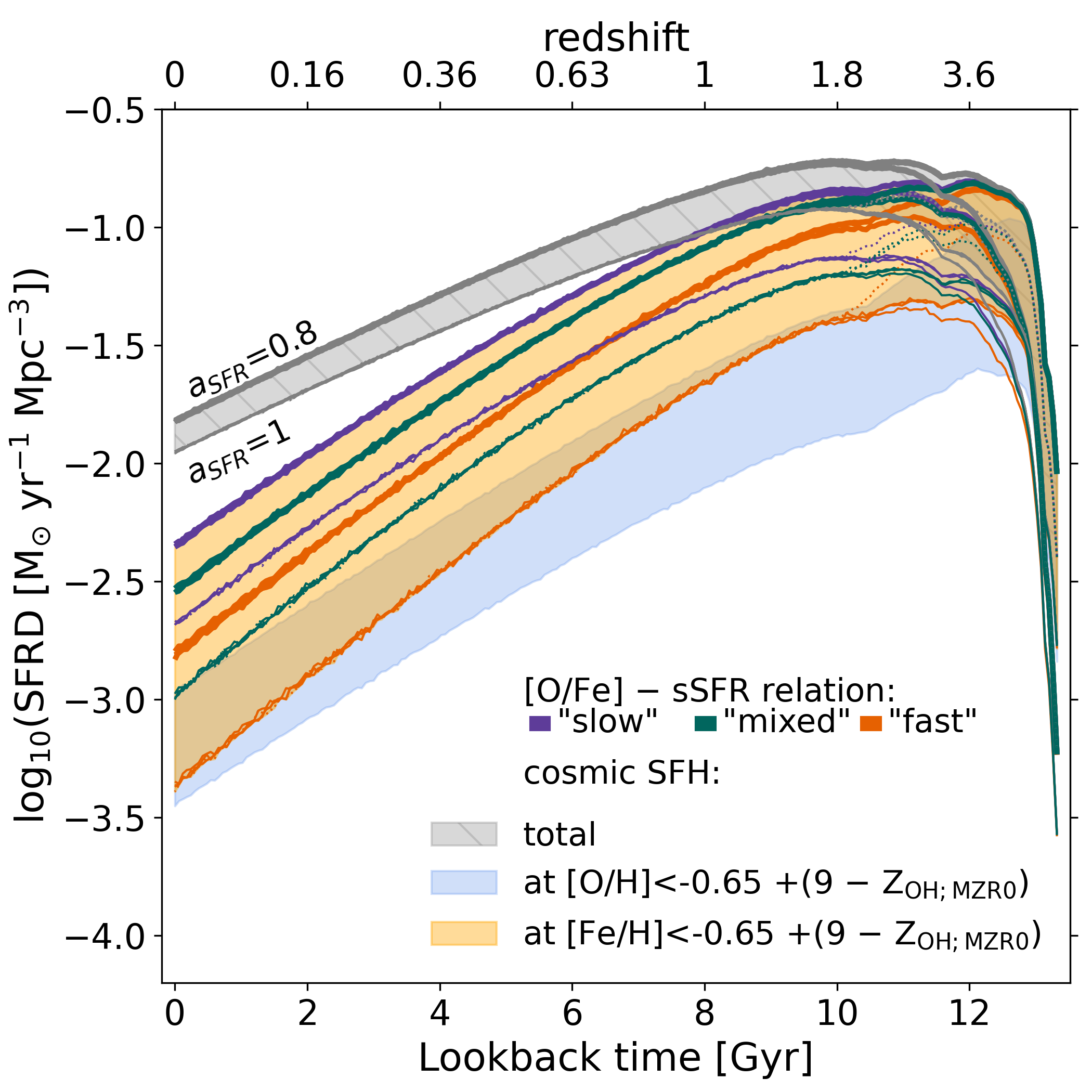}
    \hspace{0.005\textwidth}
    \includegraphics[width=0.491\textwidth]{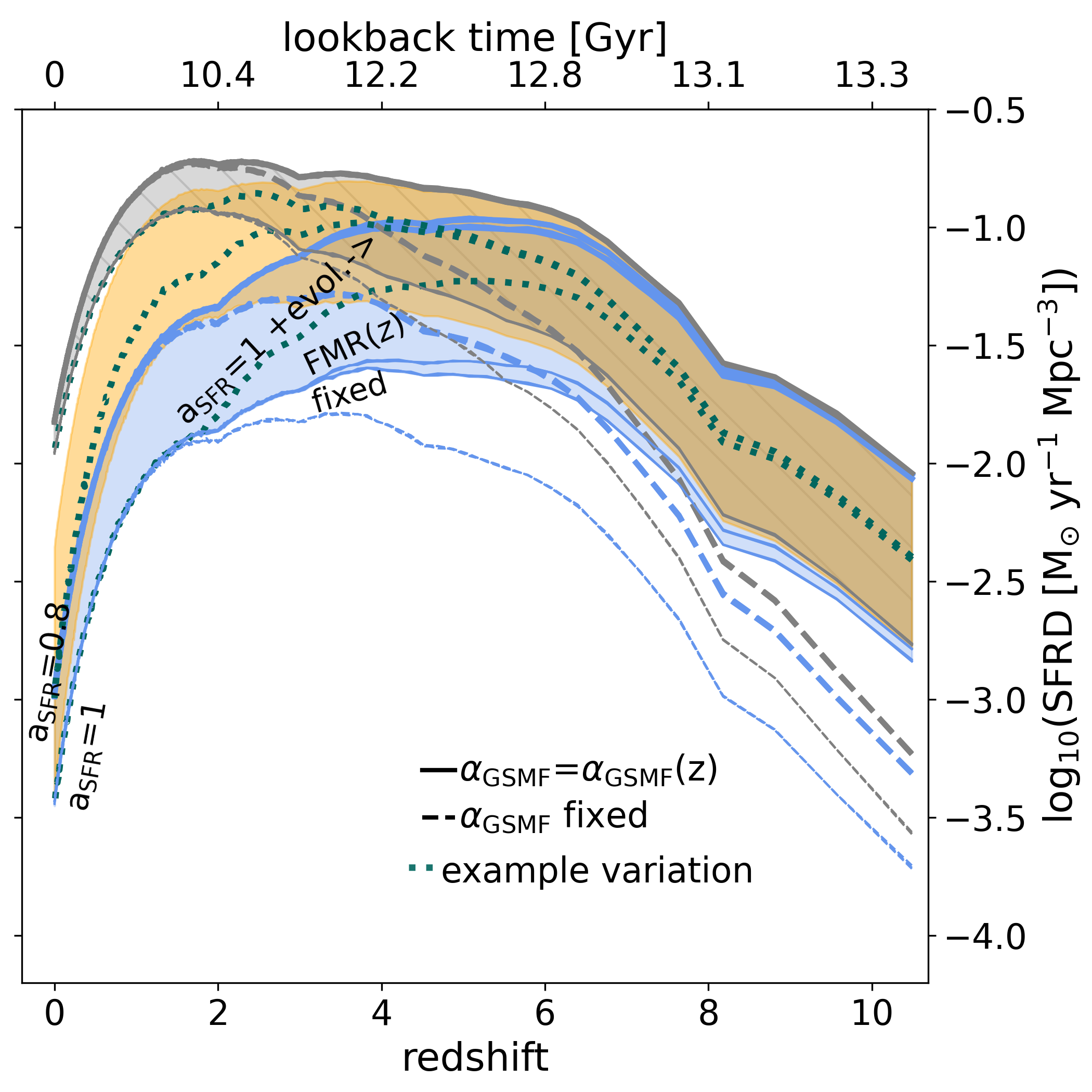}
    \caption{
    SFRD as a function of redshift or lookback time. Gray: total; orange/blue: SFRD at [Fe/H]/[O/H] below -0.65 threshold, chosen to roughly select sub-SMC metallicity environments. Shaded regions span the range of results for different model variations with $\alpha_{\rm GSMF}(z)$. 
    The left/right panels use lookback time/$z$ as the primary x-axis to show assumptions most relevant at low/high $z$. Green dotted lines in the right panel show the cSFH and its [Fe/H]<-0.65 and [O/H]<-0.65 cuts for the example variation discussed in Sec.~\ref{sec: results-example variation}. Other lines highlight selected variations, shown for iron- and oxygen-dependent cSFHs in the left and right panels, respectively.
    Systematics in the $Z_{\rm OH;MZR0}$ translate into shifts of the [O/H] and [Fe/H] thresholds, as shown in the legend.
    }
    \label{fig: lowZ csfh}
\end{figure*}
The “low-metallicity” cSFH (obtained by considering cosmic SFRD only below some metallicity threshold) is where the differences between oxygen- and iron-based results are most pronounced.
The dotted lines in the right panel of Fig. \ref{fig: lowZ csfh} illustrate this for the example variation discussed in Sec. \ref{sec: results-example variation}, and compare the total (top line) and the “low-metallicity” cSFH traced by either $Z_{\rm Fe/H}$ (middle line) or $Z_{\rm O/H}$ (bottom line).
These are obtained by integrating $f_{\rm SFR}(Z_{\rm Fe/H}, z)$ and $f_{\rm SFR}(Z_{\rm O/H}, z)$ (shown in Figs. \ref{fig: SFRD_Z_z_example} and \ref{fig: SFRD_OH_z_example}) up to the same relative “low-metallicity” threshold, [Fe/H] = [O/H] = –0.65, defined as the logarithm of the elemental abundance relative to solar.
This threshold is chosen arbitrarily, but roughly selects sub-SMC metallicity environments\footnote{Here we assume the SMC gas phase $Z_{\rm O/H}\approx$8.2 and $Z_{\rm Fe/H}$=6.85 ([Fe/H]=-0.65 on our solar scale) as in \citetalias{Chruslinska24_OFe}. However, a broad range of [Fe/H]$\approx$-1 to -0.5 were reported for young SMC stellar samples with ages of a few Myr (e.g. \citealt{Romaniello08,Narloch21}).}.
For the example variation, nearly all SFRD at $z \gtrsim 4$ happens below the SMC-like [Fe/H], whereas this occurs at $z \gtrsim 8$ if this threshold is defined in terms of oxygen abundance. The orange and blue regions shown in Fig. \ref{fig: lowZ csfh} span the cosmic SFRD below [Fe/H] < –0.65 and [O/H] < –0.65, respectively, calculated across model variations and illustrating the associated uncertainties. While the quantitative outcome depends on the chosen metallicity threshold (with lower thresholds yielding larger uncertainties) the qualitative behaviour remains consistent with that described for the example variation. 
The overlap between the two ranges increases with $z$ as the fraction of SFRD below the metallicity threshold approaches the total (gray area), occurring at lower $z$ for iron- than for oxygen-based cut. When compared for the same relative threshold, the SFRD is higher at “low” $Z_{\rm Fe/H}$ than at “low" $Z_{\rm O/H}$, and the low-$Z_{\rm Fe/H}$ cSFH peaks at lower $z$ than its oxygen counterpart.
\\
At $z\lesssim$4, the orange range is wider than the blue one due to additional uncertainty introduced by the [O/Fe] $-$ sSFR relation. 
As shown in Sec. \ref{sec: solar O/Fe}, in the "fast"/"slow" Fe-enrichment scenario (orange/purple lines in the left panel in Fig. \ref{fig: lowZ csfh}), the difference between the $f_{\rm SFR}$($Z_{Fe/H}$, $z$) and the $f_{\rm SFR}$($Z_{O/H}$, $z$) (therefore, also between low-$Z_{\rm Fe/H}$ and low-$Z_{\rm O/H}$ cSFH) is the smallest/largest.
Other assumptions, highlighted in the right panel in Fig. \ref{fig: lowZ csfh}, affect both $f_{\rm SFR}$($Z_{Fe/H}$, $z$) and $f_{\rm SFR}$($Z_{O/H}$, $z$).
The assumed non/evolution of the FMR at $z>4$ has a negligible effect on those results compared to $\alpha_{\rm GSMF}$ and a$_{\rm SFR}$.
\\ 
An alternative approach to comparing $f_{\rm SFR}(Z_{\rm Fe/H}, z)$ and $f_{\rm SFR}(Z_{\rm O/H}, z)$ (which avoids the need for a metallicity threshold) is to examine their metallicity distribution functions (MDFs), obtained by integrating the distributions over selected redshift ranges. This view can help inform the initial metallicity distribution assumptions in models of stellar or prompt-transient populations formed within specific time ranges and serve as a reference for comparison with observed or simulated galactic MDFs.
We discuss the MDFs for our models for selected redshift intervals in Appendix~\ref{sec: MDFs}.

\section{<[Fe/H]>: models and constraints from LGRBs} \label{sec: Fe/H avg. comparison}

\begin{figure*}[htbp]
    \centering
    \includegraphics[width=0.9\textwidth]{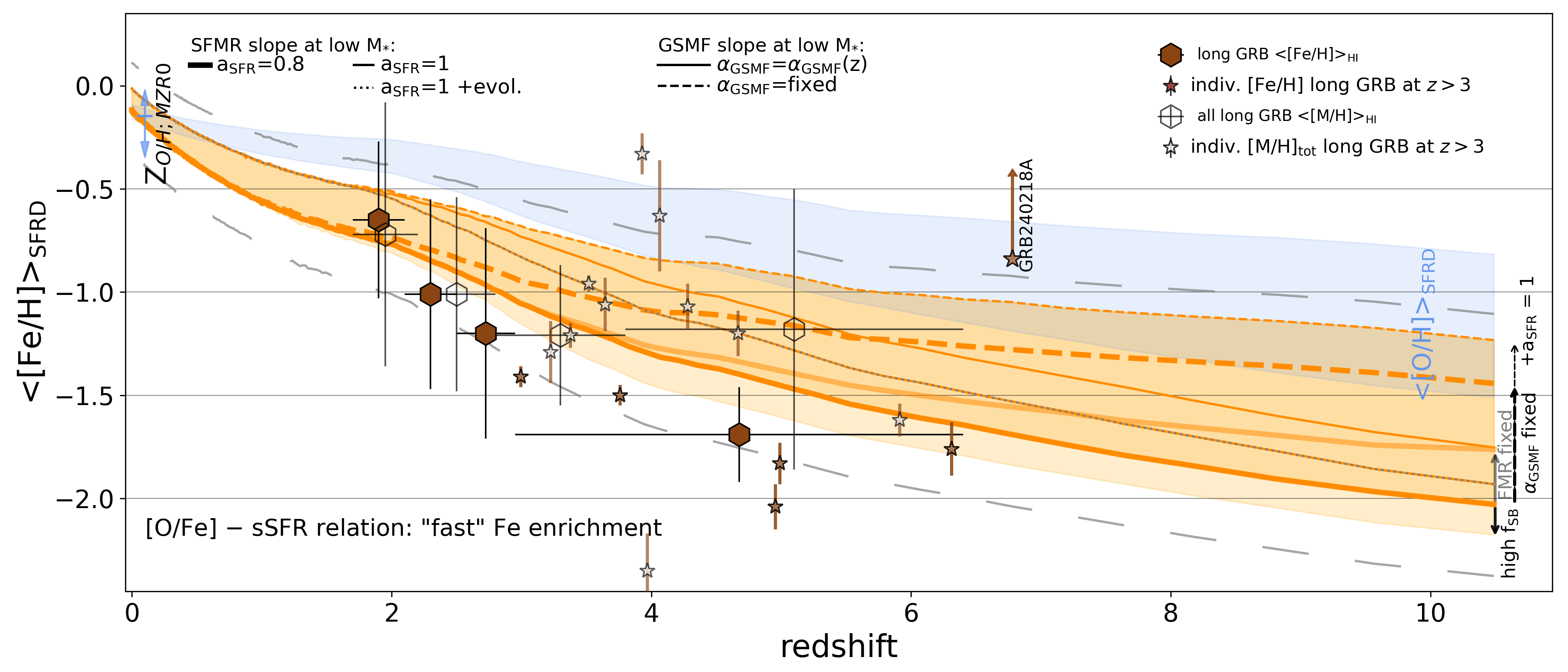}
    
    \vspace{0.4em} 
    
    \includegraphics[width=0.9\textwidth]{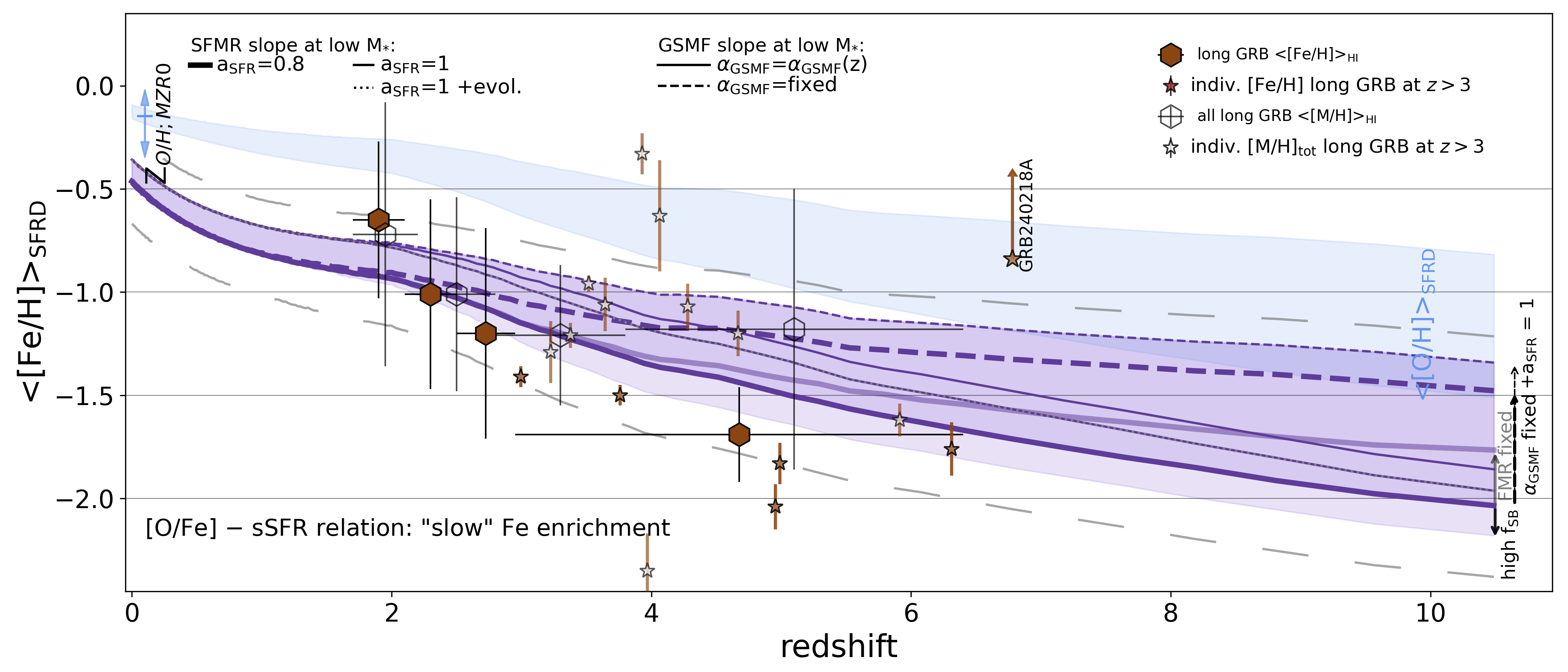}
    
    \caption{
SFRD-weighted mean $\langle \mathrm{[Fe/H]} \rangle_{\rm SFRD}$ as a function of redshift, spanned by our models (orange/purple regions) with selected variations shown as lines. The corresponding $\langle \mathrm{[O/H]} \rangle_{\rm SFRD}$ range is shown in pale blue (same in both panels). Top/bottom panels: assuming [O/Fe]–sSFR relation with “fast”/“slow” Fe enrichment. 
Data: $\langle \mathrm{[Fe/H]} \rangle_{\rm HI}$ from absorption-derived LGRB hosts (filled/open hexagons: subsample with more accurate dust corrections/all, vertical/horizontal errors - standard deviation/$z$ bin width). Stars: individual $z>3$ LGRB hosts (brown: with more accurate dust corrections for which [M/H]$_{\rm tot}\approx$[Fe/H], white: remaining objects, for which [M/H]$_{\rm tot}$ may overestimate [Fe/H]).
Vertical arrows - systematics in our estimates due to $Z_{O/H}$ calibration (affecting both $\langle \mathrm{[Fe/H]} \rangle_{\rm SFRD}$ and $\langle \mathrm{[O/H]} \rangle_{\rm SFRD}$); dashed gray lines show its effect on the edges of the $\langle \mathrm{[Fe/H]} \rangle_{\rm SFRD}$ range. %Horizontal lines every 0.5 dex are shown for reference.
}
    \label{fig: lgrb combined}
\end{figure*}

In Fig.~\ref{fig: lgrb combined}, we show $\langle \mathrm{[Fe/H]} \rangle_{\rm HI}$ derived from absorption-based LGRB host properties against $\langle \mathrm{[Fe/H]} \rangle_{\rm SFRD}$ from our model variations. 
As discussed in Sec.~\ref{sec: LGRB sample}, LGRB host sample is not selected by galaxy brightness (i.e. is not expected to be biased against faint galaxies) and additionally, LGRBs may preferentially occur in low-[Fe/H] and/or high-SFR environments.  
Consequently, $\langle \mathrm{[Fe/H]} \rangle_{\rm HI}$ could be biased to lower values relative to $\langle \mathrm{[Fe/H]} \rangle_{\rm SFRD}$ but the offset should \emph{decrease} with $z$ as LGRB-selected galaxies become increasingly representative of the general star-forming population.
In this view, models predicting $\langle \mathrm{[Fe/H]} \rangle_{\rm SFRD}$ that fall below the LGRB-based average  or diverge from it with increasing $z$ are disfavoured. The current limited sample size and the significant scatter within each $z$ bin (as indicated by the errors in $\langle \mathrm{[Fe/H]} \rangle_{\rm HI}$, representing the standard deviation) prevent us from drawing firm conclusions about any of the model variations.
Nonetheless, we note the following:
\begin{itemize}
\item Assuming a 0.2 dex lower $Z_{O/H; MZR0}$ (long-dashed gray line; within systematics in the low $z$ gas-phase $Z_{O/H}$ calibrations), would cause variations assuming "slow" Fe enrichment (bottom panel) to fall below the LGRB-based average in the lowest $z$ bin, which is not expected.
\item At $z\lesssim 3$, our estimates of $\langle \mathrm{[Fe/H]} \rangle_{\rm SFRD}$ are the most sensitive to the adopted [O/Fe]–sSFR relation. The steeper decline between $z=2$ and 3 obtained under the “fast” Fe-enrichment assumption (top panel) appears more consistent with the LGRB-based trend when taken at face value.
\item 
The underlying LGRB host population appears insufficiently sampled to allow for a reliable estimate of $\langle \mathrm{[Fe/H]} \rangle_{\rm HI}$ in the highest-$z$ bin ($z=3$–6.4). In this bin, the sub-sample of LGRB hosts selected to ensure [M/H]$_{\rm tot}\approx$[Fe/H] exhibits a much smaller scatter in [M/H]$_{\rm tot}$ compared to both other redshift bins and the full LGRB sample within the same bin. The lower limit on [Fe/H] from GRB 240218A at $z\sim6.8$ lies $\approx$0.8 dex above the estimated $\langle \mathrm{[Fe/H]} \rangle_{\rm HI}$ for $z=3$–6.4, further indicating that the small observed scatter is not physical and that the inferred $\langle \mathrm{[Fe/H]} \rangle_{\rm HI}$ may be biased.
\end{itemize}
In light of the above, we avoid drawing conclusions from the highest-$z$ bin. However, differences between our model variations become more pronounced in this regime, and future LGRB constraints would offer a valuable test for our framework.
We also show the range of $\langle \mathrm{[O/H]} \rangle_{\rm SFRD}$ spanned by our models (pale blue region in Fig. \ref{fig: lgrb combined}), which is systematically higher than both $\langle \mathrm{[Fe/H]} \rangle_{\rm SFRD}$ and $\langle \mathrm{[Fe/H]} \rangle_{\rm HI}$. 

\section{The risks of using solar-scaling}\label{sec: implications}% 
The prevalence of super-solar O/Fe abundance ratios contrasts with the fact that the solar abundance pattern is a common underlying assumption in models of stellar evolution, populations and spectra.
That assumption is inherited when such models are used to quantify the effects of stars and stellar populations on their environments and to interpret observations.
Enforcing a solar-scaled ratio where it does not apply can introduce systematic errors. 
The magnitude of these errors is inherently model-dependent, nonetheless, our results can be combined with specific modelling frameworks to quantify the impact in a given application. Here we restrict ourselves to a qualitative discussion and illustrative examples of the expected biases.

\subsection{Pedagogical example}
In models/simulations, metallicity effects are often parametrized by a simple scaling with the total metal mass fraction relative to solar ($Z/Z_\odot$). 
In practice, however, particular elements may dominate "metallicity" effects for different physical processes. 
In particular, gas cooling is largely determined by the abundance of oxygen \citep{Richings_2014,katz2022,Sharda23}, while the effects of stellar evolution (e.g. feedback, winds) are driven by the abundance of iron (see sec. \ref{sec: disc. SE}). If [O/Fe]=0, one may apply the same scaling to describe both processes, i.e. $\rm n_{O/H}/n_{O/H_\odot} = n_{Fe/H}/n_{Fe/H_\odot}=Z/Z_\odot$.
However, our results show that this is rarely the case. 
As a simple illustration of the bias introduced by this scaling, one can compare the difference in $n_{O/H}$ and $n_{Fe/H}$ when $Z/Z_\odot$ is fixed but computed for an arbitrary [O/Fe] (all the other elements are solar-scaled).
Assuming [O/Fe]=[O/Fe]$_{\rm CCSN}$=0.52 and mass fraction of $X_{O}/Z_{\odot}\approx0.5$ and $X_{Fe}/Z_{\odot}\approx0.09$ in the solar composition \citep{GrevesseSauval98}:
\\
\begin{math}
\left(\frac{\mathrm{n_{O/H}}}{\mathrm{n_{O/H}}_\odot}\right) / \left(\frac{Z}{Z_\odot}\right)
= \frac{(X_{O}+X_{Fe}) \ 10^{[\rm O/Fe]}}{X_{Fe} + X_{O} \ 10^{[\rm O/Fe]}}
\approx 1.12
\end{math}
\\
\begin{math}
\left(\frac{\mathrm{n_{Fe/H}}}{\mathrm{n_{Fe/H}}_\odot}\right) / \left( \frac{Z}{Z_\odot} \right) = \frac{(X_{O}+X_{Fe})}{X_{Fe} + X_{O}\ 10^{[\rm O/Fe]}}
 \approx 0.35
\end{math}
\\
Thus, using  $Z/Z_\odot$ scaling would overpredict $n_{O/H}$-scaled effects by only a small amount (12\%) but underpredict $n_{Fe/H}$-scaled effects by a factor 3 in young and highly star forming environments, biasing modelled cooling and feedback in opposite directions. In general, the difference is larger for elements that contribute less by mass.

\subsection{Stellar/multiple evolution} \label{sec: disc. SE}

Stellar evolution is sensitive to birth chemical composition predominantly via effect on winds, opacity and nuclear burning efficiency. 
For stars with $M_{\rm ZAMS}\gtrsim 1.2,\Msun$ whose main-sequence burning is CNO-cycle dominated, the nuclear rate depends directly on the total C+N+O abundance \citep{KippenhahnWeigert2012}. 
The main impact is on the main-sequence lifetime, shortened by a few percent if $Z_{\rm O/H}$ is increased at fixed $Z_{\rm Fe/H}$ \citep{Pietrinferni2009,VandenBerg12}. The effects of opacity and line-driven winds are more striking and increasingly important for hot, massive stars. Photospheric and envelope opacities are set largely by Fe-group abundances (\citealt{Iglesias96}; oxygen dominates only in the cool atmospheres of late-K/M dwarfs and giants) and the radiation-driven winds of hot stars scale with $Z_{\rm Fe/H}$ \citep[e.g.][]{Pauldrach93,Kudritzki02,Vink05,Puls08,Krticka18,VinkSander21,Bjorklund23}.
Mass loss is weaker at lower $Z_{\rm Fe/H}$, though the scaling also depends on stellar effective temperature, mass and luminosity. Adopting $\dot{M}_{\rm wind} \propto Z_{Fe/H}^{\alpha_{wind}}$ with $\alpha_{wind}$ = 0.42/$\alpha_{wind}$ =0.85 \citep[][above/below their bi-stability jump]{VinkSander21}, in young and strongly star-forming environments where $\mathrm{[O/Fe]}\approx\mathrm{[O/Fe]}_{\rm CCSN}$, one may overestimate $\dot{M}_{\rm wind}$ by a factor of $\sim 1.7 - 2.7$ using $Z_{\rm O/H}$ as a solar-scaled proxy for $Z_{\rm Fe/H}$.
A more striking yet uncertain example is that of very massive stars  (VMS; $\gtrsim100 \ \Msun$), which can drive optically thick winds with $\dot{M}_{\rm wind}$ that exceeds that of canonical OB stars \citep{GrafenerHamann08,Vink11,Goswami22,Sabhahit23,Costa25,Martinis25}. Such winds cause a VMS to lose most of its mass over the main sequence. However, constraints are currently lacking at sub-LMC metallicity. \citet{Sabhahit23} proposed a $Z_{\rm Fe/H}$-dependent condition for launching strong optically thick winds, which requires the VMS to be more extreme (i.e. more massive and luminous) at lower $Z_{\rm Fe/H}$. In the presence of such a threshold, incorrectly assigning a metallicity (and thus the wind mode) could affect the masses of the star and the remnant black hole (BH) by an order of magnitude. Using $Z_{\rm O/H}$ as a solar-scaled proxy for $Z_{\rm Fe/H}$ would lead to an overestimation of the strong wind phenomenon in VMS and underestimation of remnant masses - particularly at low $Z_{\rm Fe/H}$.
In summary, $Z_{\rm Fe/H}$-dependent  wind mass loss appears to be the main driver of metallicity effects on the evolution of massive stars (including in binaries/multiples). In binaries, composition-dependent radius evolution is another key factor, as it shifts when (and if) Roche-lobe overflow occurs, altering the mode, stability, and outcome of mass transfer \citep[e.g. sec 2 in][and references therein]{Costa24}. 
Radii of massive stars are uncertain, but their sensitivity to metallicity is mostly due to changes in opacity and, to a lesser extent, nuclear burning efficiency \citep[][]{Xin22}.  
In the absence of self-consistent massive star/binary models with non-solar abundance patterns representative of the considered environments, $Z_{\rm Fe/H}$ provides a practical scaling of their metallicity dependence. Conversely, relying on $Z_{\rm O/H}$ (or $\alpha$-elements) leads to largest errors.

\subsection{Galaxy/stellar population spectra and diagnostics}

Models with non-solar and solar -scaled abundances differ in their predictions for both stellar evolution (isochrones, evolutionary tracks) and atmospheres. 
In models of old populations (which formed when <[O/Fe]>$_{\rm SFRD}$ was high, e.g. ellipticals, GCs),
the key differences lie in colours, particularly for turnoff/subgiant branch stars used as age indicators, leading to biased age determinations if ignored \citep[e.g.][]{Trager00,ThomasMaraston03,Vazdekis15,Coelho14,Pietriferini21,Knowles23,Park24}.
Large grids of $\alpha$-enhanced evolutionary models of (very) massive stars, which provide the basis for self-consistent SPS models suitable for interpreting young, highly star-forming environments typical of high-$z$ galaxies, have only recently begun to appear in the literature \citep[][]{Grasha21,Byrne22,Byrne25}. In massive-star dominated populations, the key differences lie in the UV continuum, ionizing outputs, and UV wind features.
These are primarily driven by $Z_{\rm Fe/H}$ rather than non-solar [O/Fe] \emph{per se} (see sec \ref{sec: disc. SE}).
In general, a young stellar population at lower $Z_{\rm Fe/H}$ is more UV-bright and produces a harder ionising spectrum \citep[e.g.][]{Leitherer99,SmithNorrisCrowther2002, Stanway16,StanwayEldridge18}.
Using $Z_{\rm O/H}$ as a proxy for $Z_{\rm Fe/H}$ when deriving SFR with common UV diagnostics \citep[e.g.][]{BickerFritze2005,HaoKennicuttCalzetti2011,KennicuttEvans2012,MadauDickinson14} would lead to an overestimation in environments with [O/Fe]>0.
Furthermore, ignoring the fact that in such environments iron-poor stars are ionizing relatively oxygen-rich nebulae can lead to misinterpretation of stellar and nebular diagnostics \citep[e.g.][]{Steidel16,Trainor16,Strom17}.
As gas  cooling and heating are determined by $Z_{\rm O/H}$ and $Z_{\rm Fe/H}$, respectively, this can be effectively accounted for in the models by adopting different metallicities for the stellar and gas components \citep{Strom18,Sanders20,Runco21,Strom22}.

 \subsection{ Transients and phenomena linked to metal-poor progenitors }
 \begin{figure}[htbp]
    \centering
    \includegraphics[width=\columnwidth]{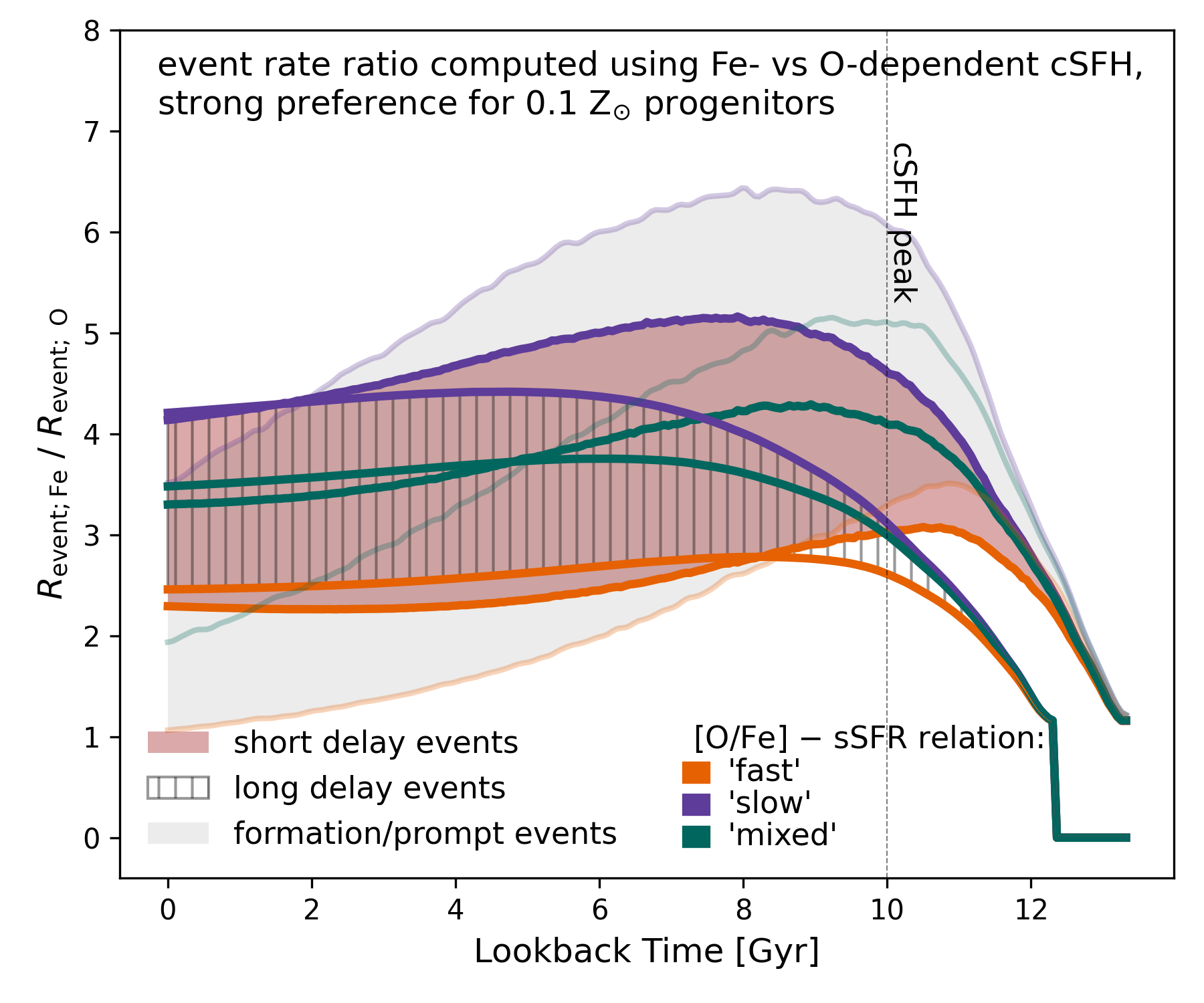}
    
    \caption{
Relative rate density of events originating from metal-poor progenitors, obtained using $f_{\rm SFR}(Z_{\rm Fe/H},t)$ and $f_{\rm SFR}(Z_{\rm O/H},t)$ as a function of lookback time.
Gray area - prompt events/formation rate, brown/hatched area - events with a distribution of delays, peaked at short (10 Myr)/long (1 Gyr) delay times with respect to star formation.
coloured lines distinguish assumptions about the [O/Fe]-sSFR relation, other assumptions relevant for the $f_{\rm SFR}(Z_{\rm Fe/H},t)$  and $f_{\rm SFR}(Z_{\rm O/H},t)$  are the same as for the example variation from Sec. \ref{sec: results-example variation}.
Vertical line indicates the time of the peak of the cSFH.
Using $f_{\rm SFR}(Z_{O/H},t)$ underestimates metal-poor event rates by a redshift (and model) dependent factor.
}
    \label{fig: metal poor rates}
\end{figure}
The uncertainty and difference between Z$_{\rm Fe/H}$ and Z$_{\rm O/H}$ -dependent cSFH is the largest in the metal-poor regime.
This is also where many transients of stellar-origin appear more common and/or more luminous. 
These include X-ray sources \citep{Kovlakas20,Fornasini20,Lehmer21},  LGRBs and hydrogen-poor superluminous supernovae \citep{Lunnan14,Schulze18} or luminous fast blue optical transients \citep{Coppejans20,Chrimes24,Chrimes24a,Klencki25}.
Phenomena theorized to prefer metal-poor progenitors also include stellar BH mergers \citep{Belczynski10,Belczynski16,Klencki18,Giacobbo18,ChruslinskaNelemansBelczynski19,Neijssel19,Broekgaarden22,Iorio23,Chruslinska24}, pair-instability supernovae \citep{Woosley02,Stevenson19,Sabhahit23}, and chemically homogeneous evolution \citep{YoonLanger05,YoonLangerNorman06,Szecsi15,Marchant16,MandelDeMink16}. 
Stellar BHs with masses $\gtrsim 30 \Msun$, which are found in a substantial fraction ($\gtrsim$40\%) BH merger events detected with GW \citep{GWTC-4}, may also require subsolar progenitors \citep{Belczynski10BHmax,Romagnolo23,Vink24,Merritt25}.
Furthermore, harder ionising spectra than predicted by existing SPS models are required to explain the emission properties of metal-poor galaxies \citep{Senchyna17,Schaerer19,Katz23,Telford23}.
Overall, it appears that stellar evolution is more likely to produce (more of the) extreme phenomena at lower metallicities. However, the driving factors and the regime in which these can occur remain unclear. As in the above examples, most of the empirical constraints come from stellar/transient populations. These require metallicity-dependent cSFH as part of the modelling and interpretation.\\
In Figure \ref{fig: metal poor rates} we present a simple estimate of the relative occurrence rate of events with strong preference for low metallicity progenitors calculated using $f_{\rm SFR}(Z_{\rm Fe/H},t)$ and $f_{\rm SFR}(Z_{\rm O/H},t)$. Assuming that in the context of stellar-origin transients and phenomena "metallicity"$\approx Z_{\rm Fe/H}$, this serves to show the bias introduced by the use of metallicity-dependent cSFH based on an incorrect probe. We note that the result shown in Fig. \ref{fig: metal poor rates} is model-dependent and it should be taken as a \emph{sign and order-of-magnitude} calculation.
The event rate $R_{\rm event; X}$ follows from:
\\
\begin{equation}
R_{\rm event; X}(t_0)
= \int_{0}^{t_0} \int
f_{\rm SFR}(Z_{X/H},\, t')\;
\eta(Z_{X/H})\;
p_{\rm delay}(t_0 - t')\;
\, dZ_{X/H}\, dt'
\label{eq:event_rate_integral}
\end{equation}
where index 'X' denotes either oxygen (X=O) or iron (X=Fe). $\eta(Z_{X/H})$ stands for the event progenitor formation efficiency (number per unit stellar mass formed) as a function of metallicity, for which we use a logistic function \citep[used in the context of LGRBs, hydrogen-poor superluminous SNe, BH mergers, c.f.][]{Schulze18,Chruslinska24,Disberg25}:
\begin{equation}
\eta(Z_{X/H}) =
       \eta_{high} + (\eta_{low}-\eta_{high})/ (1. + \mathrm{exp}(-k\cdot(Z_{X/H} - Z_{X/H_{thr}})))
\label{eq:eta_logistic}
\end{equation}
with $k=-10$, which leads to a sharp decrease from $\eta_{low}$ to $\eta_{high}$ with a midpoint at $Z_{X/H_{thr}}$ and we set this midpoint value to 1/10 of the solar oxygen/iron abundance.
We further assume $\eta_{high}=\eta_{low}/100$, i.e. the formation efficiency is not fully truncated at high $Z_{X/H}$ but drops by two orders of magnitude with respect to plateau $\eta_{low}$ at low $Z_{X/H}$.
Since we only compare the relative rate, $\eta_{low}$ value is irrelevant.
For the delay time distribution, we assume the power-law 
\begin{equation}
p_{\rm delay}(\tau) =
\begin{cases}
\propto \tau^{-1}, & \tau \ge \tau_{\rm event;min},\\[3pt]
0, & \tau < \tau_{\rm event;min},
\end{cases}
\label{eq:dtd_powerlaw}
\end{equation}
normalised to unity between $\tau_{\rm event;min}$ and Hubble time. This is a conventional approximation for $p_{\rm delay}$ of gravitational wave sources (e.g. BH mergers).
We use $\tau_{\rm event;min}$=10 Myr and 1 Gyr for events with a "short" (brown region) and "long" delay (hatched region), respectively, for the examples shown in Fig. \ref{fig: metal poor rates}.
Both short and long delays are possible for BH mergers within the formation scenarios considered in the literature.
Replacing the $p_{\rm delay}$ with a Dirac delta at zero, equation \ref{eq:event_rate_integral} can be used to compute the formation rate/prompt event rate (e.g. LGRB, supernovae).
We point out that:
\begin{itemize}
\item Using $f_{\rm SFR}(Z_{O/H},t)$ underestimates metal-poor event rates. The difference varies with $z$/lookback time.
\item The difference depends on $f_{\rm SFR}(Z_{Fe/H},t)$ model variation and is the largest for variations with the largest difference between O- and Fe- dependent cSFH (in particular, for 'slow' Fe enrichment model family).
\item  The difference also depends on $p_{\rm delay}$ (larger for prompt events),  $\eta(Z_{X/H})$ (larger for more extreme metallicity threshold/dependence).
\item $f_{\rm SFR}(Z_{Fe/H},t)$ model variation  affects the metal-poor event rate evolution over cosmic time/redshift. In particular, uncertainty in the [O/Fe]-sSFR relation alone affects the evolution, peak location and the magnitude of the rate (factor of 2-3).
\end{itemize}
Variations in $R_{\rm event}$ due to uncertainty in $f_{\rm SFR}(Z_{Fe/H},t)$ across our variations are typically within a factor of a few but can reach $\sim O(10)$ for more extreme $\eta(Z_{Fe/H})$ and $p_{\rm delay}$ choices than used in our example.
The bias introduced by the use of $f_{\rm SFR}(Z_{O/H},t)$ and uncertainty of the $f_{\rm SFR}(Z_{Fe/H},t)$ are both important when considering transients/phenomena sensitive to low metallicity.

\section{Conclusions}
Oxygen and iron are key to different astrophysical processes, and are primarily produced by nucleosynthetic sources operating on different timescales. Their relative abundance in the interstellar medium varies over time and can significantly depart from the solar ratio.
However, observationally, it is often only possible to constrain one of these two elemental abundances, which is then used as a representative proxy for metallicity after being scaled to the solar abundance.
Usually, the different roles and evolution of these elements are not distinguished in models and simulations that use a total metal mass fraction to parametrise the effects of metallicity on different processes.
In this work, we propose a framework to overcome these observational limitations and provide a more appropriate scaling for the cosmic evolution of the abundances of oxygen and iron that can be used in models.
\\
Building on Paper I \citetalias{Chruslinska24_OFe}, which introduced the theoretical underpinnings of the redshift-invariant [O/Fe]–sSFR relation of galaxies and constrained it empirically, we apply that relation to derive the iron-dependent cosmic star formation history, $f_{\rm SFR}(Z_{\rm Fe/H},t)$, alongside its oxygen-based counterpart, $f_{\rm SFR}(Z_{\rm O/H},t)$.
Here, we extend and validate the observation-based framework of \citetalias{ChruslinskaNelemans19,Chruslinska21} against more recent and additional empirical constraints.
 We also introduce several model variations to bracket the main uncertainties in the core ingredients of our framework, i.e. the distributions of galaxy SFR , $Z_{O/H}$, GSMF and [O/Fe]-sSFR relation. 
 This enables us to quantify how these uncertainties propagate to the final result. Notably, the high-$z$ JWST results constrain the evolution of the FMR and MZR at $z>3$ enough so that it is no longer the dominant source of uncertainty in our $f_{\rm SFR}(Z_{\rm O/H},t)$ estimates.
Our main conclusions are the following: 
\begin{itemize}
\item Star formation with a near-solar [O/Fe] abundance is rare throughout cosmic history; the bulk (at least $70\%$ across our variations) of the integrated stellar mass formed with O/Fe larger than the solar abundance ratio. 
\item The SFRD-weighted average metallicity is lower when using [Fe/H] than [O/H] at all redshifts, except when the [O/Fe]–sSFR relation resembles our “fast Fe-enrichment” case, for which $\langle \mathrm{[O/Fe]} \rangle_{\rm SFRD}$ may become subsolar at $z\lesssim0.4$ (Fig. \ref{fig: OFe_avg}). 
\item The offset between $\langle \mathrm{[Fe/H]} \rangle_{\rm SFRD}$ and $\langle \mathrm{[O/H]} \rangle_{\rm SFRD}$ grows to $z\approx 3$ and approaches a roughly constant value at higher redshift  ([O/Fe]$\approx$0.4-0.5 dex, i.e. up to the CCSN-ratio; Fig. \ref{fig: OFe_avg}). The offset depends primarily on the [O/Fe]–sSFR relation (particularly on the assumed [O/Fe]$_{\rm CCSN}$ for core-collapse supernovae and the minimum delay for enhanced Fe enrichment) and on the galaxy SFR distribution.
\item The differences between $f_{\rm SFR}(Z_{\rm O/H},t)$ and $f_{\rm SFR}(Z_{\rm Fe/H},t)$ are the largest in the low-metallicity and high-$z$ regimes. These differences are the smallest if the delay in the enhanced Fe enrichment relative to star formation is minimal (our “fast Fe-enrichment” case; Figs. \ref{fig: SFRD_Z_z_example} and \ref{fig: lowZ csfh}).
\item The absorption-derived $\langle \mathrm{[Fe/H]} \rangle_{\rm HI}(z)$ from LGRB afterglow–selected sources is broadly consistent with our $\langle \mathrm{[Fe/H]} \rangle_{\rm SFRD}(z)$ estimates and is offset from $\langle \mathrm{[O/H]} \rangle_{\rm SFRD}(z)$ (Fig. \ref{fig: lgrb combined}). While the current comparison is not constraining for our models, a larger sample (particularly at $z>4$) could provide the most direct tests of our $f_{\rm SFR}(\mathrm{Z_{Fe/H}},t)$ framework.
\end{itemize}
These results underscore that “metallicity” estimates based on oxygen  (or $\alpha$-elements in general) and iron-group probes are not interchangeable. Failing to account for non-solar abundance ratios may lead to important biases and systematic errors.
In particular, in environments with [O/Fe] > 0:
\begin{itemize}
\item Relying on solar abundance scaling would slightly overpredict O-scaled effects (e.g. gas cooling) and underpredict Fe-scaled effects (e.g. stellar feedback; by up to a factor of $\sim3$).
\item Using $Z_{\rm O/H}$ as a solar-scaled proxy for $Z_{\rm Fe/H}$ overestimates hot-star/VMS wind mass loss effects. This non-trivially affects the evolution of stars/multiples, UV and ionising stellar spectra (SFR and metallicity diagnostics), properties of stellar BHs and transients of massive-star origin.
\item Using $f_{\rm SFR}(Z_{O/H},t)$ instead of $f_{\rm SFR}(Z_{Fe/H},t)$ underestimates the rates of events and phenomena linked to metal-poor progenitors (e.g. LGRBs, stellar BH mergers) by a $z$-dependent factor, complicating the interpretation of $z$-trends in their population properties (Fig. \ref{fig: metal poor rates}).
\end{itemize}
The magnitude of these effects is model-dependent. Our results can be used within specific frameworks to alleviate and quantify these biases.
Finally, we emphasise that we obtain the above results simultaneously predicting that galaxies with MW–like  $M_{*}$ are among the dominant contributors to the cosmic SFRD at $z\lesssim2$ and can form stars with the Sun's $Z_{\rm O/H}$ and $Z_{\rm Fe/H}$ at solar birth epoch. However, for model variations with long ($\gtrsim10^{8}{\rm yr}$) delays in enhanced Fe enrichment (our “slow Fe-enrichment” case) such stars would not coincide with the MDF peak in MW-like galaxies.

\section{Data availability}
Data generated during this study and code underlying our analysis are available via Zenodo \url{https://doi.org/10.5281/zenodo.18018782} and GitHub \url{https://github.com/Mchruslinska/MetalCosfr_public.git}.

\begin{acknowledgements} 
MCh would like to thank T. Matsuno and J. Klencki for their insightful comments, and K. Heintz for the discussion on the LGRB sample. We thank the anonymous referee for the comments and suggestions that have helped us to improve the paper.
\\
Software: matplotlib \citep{matplotlib}, scipy \citep{scipy}, numpy \citep{numpy}, astropy \citep{Astropy1,Astropy2}.
\end{acknowledgements}

\bibliographystyle{aa}
\bibliography{bibliography_aa.bib}

\begin{appendix}
\nolinenumbers
\section{Metallicity-dependent cosmic star formation history from galaxy scaling relations}
\label{app: method}

We model all distributions as (log)normal
\begin{equation}
    \mathcal{N}(x;\mu,\sigma)
    =
    \frac{1}{\sqrt{2\pi}\sigma}
    \exp\!\left[-\frac{(x-\mu)^2}{2\sigma^2}\right].
\end{equation}
 where $\mu$ represents the mean and $\sigma$ dispersion.\\
At fixed galaxy stellar mass \(M_\ast\) and redshift \(z\), the SFR distribution is centered on the SFMR (sec. \ref{app: SFMR}):
\begin{equation}
    P(\log_{10}\mathrm{SFR}\,|\,M_\ast,z)
    =
    \mathcal{N}\!\left(
        \log_{10}\mathrm{SFR}_{\mathrm{SFMR}}(M_\ast,z),
        \sigma_{\mathrm{SFR}}
    \right).
\end{equation}
with $\sigma_{\mathrm{SFR}}=0.3$ dex.
In presence of the SB sequence (sec. \ref{sec, app: SB}), it becomes bimodal:
\begin{equation}
\begin{split}
    P(\log_{10}\mathrm{SFR}\,|\,M_\ast,z)
    =\left[1-f_{\mathrm{SB}}(M_\ast,z)\right]\\
   \times \mathcal{N}\!\left(
        \log_{10}\mathrm{SFR}_{\mathrm{SFMR}}(M_\ast,z),
        \sigma_{\mathrm{SFR}}
    \right)\\
    +f_{\mathrm{SB}}(M_\ast,z)
    \mathcal{N}\!\left(
        \log_{10}\mathrm{SFR}_{\mathrm{SB}}(M_\ast,z),
        \sigma_{\mathrm{SB}}
    \right),
    \end{split}
\label{eq:notes_sfr_mixture_step3}
\end{equation} with $\sigma_{SB}$=0.2 dex and the relative SB and SFMR occupation governed by the starburst fraction $f_{SB}$.
Oxygen abundance distribution is centered on the value assigned using the FMR (sec \ref{app: FMR}):
\begin{equation}
    P(Z_{\mathrm{O/H}}|M_\ast,\mathrm{SFR},z)
    =
    \mathcal{N}\!\left(
        Z_{\mathrm{O/H;FMR}}(M_\ast,\mathrm{SFR},z),
        \sigma_{\mathrm{FMR}}
    \right).
\end{equation}
with $\sigma_{\mathrm{FMR}}$=0.05 dex. 
The oxygen-based metallicity-dependent cSFH can be written as
\begin{equation}
\begin{split}
    f_{\mathrm{SFR}}(Z_{\mathrm{O/H}},z)
    =
    \int d\log_{10}M_\ast\;
    \dfrac{dn_{\text{gal}}}{d\log_{10} M_{*\text{}}}(z) \\
    \int d\log_{10}\mathrm{SFR}\;
    P(\log_{10}\mathrm{SFR}\,|\,M_\ast,z)\;
    \mathrm{SFR} \\
    \times
    P_{\nabla}\!\left(
        Z_{\mathrm{O/H}}
        \,\middle|\,
        Z_{\mathrm{O/H;FMR}}(M_\ast,\mathrm{SFR},z)
    \right),
\end{split}
\label{eq:notes_full_csfh_oh}
\end{equation}
where $\dfrac{dn_{\text{gal}}}{d\log_{10} M_{*\text{}}}$ follows from the GSMF (section \ref{sec, app: GSMF}) and 
\begin{equation}
    P_{\nabla}\!\left(
        Z_{\mathrm{O/H}}
        \,\middle|\,
        Z_{\mathrm{O/H;FMR}}
    \right)
    =
    \mathcal{N}\!\left(
        Z_{\mathrm{O/H}},
        Z_{\mathrm{O/H;FMR}},
        \sigma_{\nabla\mathrm{O/H}}
    \right)
\end{equation}
represents additional spread in star formation over metallicity inside galaxies, modelled as a normal distribution in \(Z_{\mathrm{O/H}}\) centered on the FMR value.\\
Analogously, the iron-based cSFH can be written as
\begin{equation}
\begin{split}
    f_{\mathrm{SFR}}(Z_{\mathrm{Fe/H}},z)
    =
\int d\log_{10}M_\ast\;
    \dfrac{dn_{\text{gal}}}{d\log_{10} M_{*\text{}}}(z) \\
    \int d\log_{10}\mathrm{SFR}\;
    P(\log_{10}\mathrm{SFR}\,|\,M_\ast,z)\;
    \mathrm{SFR} \\
    \times
    P_{\nabla}\!\left(
        Z_{\mathrm{Fe/H}}
        \,\middle|\,
        Z_{\mathrm{Fe/H,avg}}
    \right),
\end{split}
\label{eq:notes_full_csfh_feh}
\end{equation}
where
\begin{equation}\label{eq: avg Fe/H}
    Z_{\mathrm{Fe/H,avg}}
    =
    Z_{\mathrm{O/H,FMR}}
    -
    [\mathrm{O/Fe}]_{\rm avg}
    -
    \log_{10}(\mathrm{O/Fe})_\odot,
\end{equation}
and \([\mathrm{O/Fe}]_{\rm avg}\) follows from the \([\mathrm{O/Fe}]\)--sSFR relation, 
and
\begin{equation}
    P_{\nabla}\!\left(
        Z_{\mathrm{Fe/H}}
        \,\middle|\,
        Z_{\mathrm{Fe/H,avg}}
    \right)
    =
    \mathcal{N}\!\left(
        Z_{\mathrm{Fe/H}},
        Z_{\mathrm{Fe/H;avg}},
        \sigma_{\nabla\mathrm{Fe/H}}
    \right)
\end{equation}
Similarly, one can compute the cSFH as a function of the oxygen over iron abundance ratio:
\begin{equation}
\begin{split}
    f_{\mathrm{SFR}}(\mathrm{[O/Fe]},z)
    =
\int d\log_{10}M_\ast\;
    \dfrac{dn_{\text{gal}}}{d\log_{10} M_{*\text{}}}(z) \\
    \int d\log_{10}\mathrm{SFR}\;
    P(\log_{10}\mathrm{SFR}\,|\,M_\ast,z)\;
    \mathrm{SFR} \\
    \times
    P\!\left(
        \mathrm{[O/Fe]}
        \,\middle|\,
        \mathrm{[O/Fe]}_{avg}
    \right),
\end{split}
\label{eq:notes_full_csfh_ofe}
\end{equation}
and we introduce additional scatter in [O/Fe] to account for possible inhomogeneities in local O and Fe enrichment within galaxies:
\begin{equation}
P\!\left(
        \mathrm{[O/Fe]}
        \,\middle|\,
        \mathrm{[O/Fe]}_{avg}
    \right)
    =
    \mathcal{N}\!\left(
        \mathrm{[O/Fe]},
        \mathrm{[O/Fe]_{avg}},
        \sigma_{\mathrm{O/Fe}}
    \right)
\end{equation}
We assume $\sigma_{\nabla\mathrm{O/H}}=\sigma_{\nabla\mathrm{Fe/H}}$=0.14 dex and $\sigma_{\mathrm{O/Fe}}=0.25 \cdot \sigma_{\nabla\mathrm{O/H}}$
(see Sec. \ref{app: scatter}).

The cosmic star formation rate density within a given metallicity interval $\Delta Z_{X} = Z_{X;\max} - Z_{X;\min}$ is then given by
\begin{equation}
\mathrm{SFRD}(\Delta Z_{X}, z) = \int_{Z_{X;\min}}^{Z_{X;\max}} f_{\mathrm{SFR}}(Z_{X}, z), dZ_{X}.
\end{equation}
Here, $Z_{X}$ denotes any of the metallicity tracers considered above (oxygen abundance, iron abundance, or [O/Fe]). The total cosmic star formation history is recovered by integrating over the full metallicity range.
 In our numerical framework, \ref{eq:notes_full_csfh_oh}, \ref{eq:notes_full_csfh_feh} and \ref{eq:notes_full_csfh_ofe}
integrals are discretized and evaluated over grids in mass, metallicity (200 equally spaced bins in $Z_{O/H}$ between 5.3 and 9.7, in [Fe/H] between -4.3 and 2.3, in [O/Fe] between -0.3 and 0.8) and cosmic time $t(z \leq 10$) in 60 Myr intervals, where we sum contributions from galaxies with $\log_{10}M_\ast \in [6,12]$. 
Stellar mass density formed outside the metallicity grid is assigned to the nearest edge bin.

\subsection{Galaxy scaling relations}
Galaxy scaling relations (SFMR, MZR, FMR) are described with a common functional form:
\begin{equation}
    \log Y
    =
    \log Y_{\mathrm{asym}}(x)
    -
    B(x)\log_{10}\!\left[
        1+\left(\frac{M_\ast}{M_0(x)}\right)^{-a/B(x)}
    \right].
\end{equation}
where  \(a\) is the low-mass slope, \(M_0\) is the turnover mass, \(\log Y_{\mathrm{asym}}\) is the asymptotic high-mass value, \(B\) controls the width of the turnover and $x$ denotes a potential secondary dependence on redshift/cosmic time or SFR. 
\subsection{Star formation - mass relation}\label{app: SFMR}
The SFMR follows \citetalias{Popesso23}, Eq.~(14)--(15):
\begin{equation}
    \log_{10}\mathrm{SFR}_{\rm SFMR}(M_\ast,t)
    =
    a_0 + a_1 t_{Gyr}
    -
    \log_{10}\!\left[
        1+\left(\frac{M_\ast}{10^{a_2+a_3 t_{Gyr}}}\right)^{-a_{SFR}}
    \right].
\end{equation}
where $t_{Gyr}$ is cosmic time in Gyr.
The specific SFR of main sequence galaxies is then obtained as $\log_{10}\mathrm{sSFR}_{\rm SFMR}(M_\ast,t)=\log_{10}\mathrm{SFR}_{\rm SFMR}(M_\ast,t)-\log_{10}\mathrm{(M_\ast)}$.
The SFMR parameters are
\begin{equation}
    a_0=2.71,\quad
    a_1=-0.186,\quad
    a_2=10.86,\quad
    a_3=-0.0729,\quad
\end{equation}
and slope $a_{SFR}$=1 in \citetalias{Popesso23}. We consider an additional variation with $a_{SFR}$=0.8. We also consider a variation where  coefficient that governs the  dependence of the normalisation is set to  $a_1$=-0.3 at $z\geqslant$1.8, leading to increased galaxy SFR at high $z$. 
\subsection{Mass-metallicity relation} 
The MZR:
\begin{equation}
    Z_{\mathrm{O/H,MZR}}
    =
    Z_{\mathrm{O/H};\mathrm{MZR0}}
    -
    B_{\rm MZR}
    \log_{10}\!\left[
        1+\left(\frac{M_\ast}{M_{0;\mathrm{MZR0}}}\right)^{-a_{\mathrm{MZR}}/B_{\rm MZR}}
    \right].
\end{equation}
 is only evaluated at $z=0$ with parameters :
\begin{equation}
    a_{\mathrm{MZR}}=0.28,\quad
    \log_{10}M_{0;\mathrm{MZR0}}=10.02, \quad B_{\rm MZR}=0.23
\end{equation}
as in \citet{Curti20} and $Z_{\mathrm{O/H};0}$=9 (shifted by \(+0.2\) dex with respect to \citealt{Curti20}).
\subsection{Fundamental metallicity relation}\label{app: FMR}
The FMR follows phenomenological formulation from \citetalias{Chruslinska21} (Eq.~1 - 6 therein, recalled below), and is parametrised as in \citet{Curti20}:
\begin{equation}
    Z_{\mathrm{O/H;FMR}}(\mathrm{SFR},M_\ast)
    =
    Z_{\mathrm{O/H};0}
    -
    \frac{\gamma_{FMR}}{\beta}
    \log_{10}\!\left[
        1+\left(\frac{M_\ast}{M_{0;\mathrm{SFR}}}\right)^{-\beta}
    \right],
\end{equation}
\citetalias{Chruslinska21} defines the FMR parameters in terms of the $z=0$ SFMR and MZR parameters and additional parameter $\nabla_{\mathrm{FMR0}}$, which characterises the strength of the SFR - $Z_{O/H}$ correlation in the low $M_{*}$ regime as
\begin{equation}
    \delta_{\mathrm{MZR}} = -\nabla_{\mathrm{FMR0}}\,\delta_{\mathrm{SFMR}}.
\end{equation}
 where
\begin{equation}
    \delta_{\mathrm{MZR}} = Z_{\mathrm{O/H,gal}}-Z_{\mathrm{O/H,MZR}},
\end{equation} and
\begin{equation}
    \delta_{\mathrm{SFMR}}=\log_{10}(\mathrm{SFR}_{\mathrm{gal}})-\log_{10}(\mathrm{SFR}_{\mathrm{SFMR}}).
\end{equation}
are the galaxy's offsets from the MZR and SFMR. The slope of the linear part of the FMR is defined as
\begin{equation}
    \gamma_{\mathrm{FMR}}=a_{\mathrm{MZR}}+\nabla_{\mathrm{FMR0}}\,a_{\mathrm{SFR}}.
\end{equation}
and the turnover mass depends on galaxy SFR
\begin{equation}
\begin{split}
    \log_{10}M_{0;\mathrm{FMR}}(\log_{10}\mathrm{SFR})
    =
    \frac{\nabla_{\mathrm{FMR0}}}{\gamma_{\mathrm{FMR}}}\log_{10}\mathrm{SFR}
    \\-\frac{\nabla_{\mathrm{FMR0}}\,b_{\mathrm{SFMR}}}{\gamma_{\mathrm{FMR}}}
    +\frac{a_{\mathrm{MZR}}}{\gamma_{\mathrm{FMR}}}\log_{10}M_{0;\mathrm{MZR0}}.
    \end{split}
\end{equation}
where $b_{\mathrm{SFR}}$ is obtained from the linear low-mass part of the $z=0$ SFMR,
\begin{equation}
    \log_{10}\mathrm{SFR}=a_{\mathrm{SFR}}\log_{10}M_\ast+b_{\mathrm{SFR}},
\end{equation}
For the SFMR used in this study $b_{\mathrm{SFR}}$ follows from the P23 relation evaluated at $t_{0}$=$t_{Gyr}(z=0)$
\begin{equation}
   \log_{10}\mathrm{SFR}
    \simeq  a_{SFR} \log_{10}M_\ast
    + \bigl[2.71-0.186\,t_0-(10.86-0.0729\,t_0)\bigr],
\end{equation}
The FMR knee parameter \(\beta(SFR)\) is evaluated numerically by enforcing that the FMR reproduces the \(z=0\) MZR for galaxies on the \(z=0\) SFMR.
In the model variation with evolving FMR-normalisation we assume:
\begin{equation}
    Z_{\mathrm{O/H};0}(z>3)= -0.0357\,(z-3)+Z_{\mathrm{O/H};\mathrm{MZR0}},
\end{equation}
 and use $Z_{\mathrm{O/H};0} =Z_{\mathrm{O/H};\mathrm{MZR0}}$ otherwise.

\subsection{Galaxy Stellar Mass Function} \label{sec, app: GSMF}

The GSMF is obtained by averaging literature results grouped in several redshift bins, as detailed in \citetalias{ChruslinskaNelemans19}.
 In addition to those listed in table 1 in \citetalias{ChruslinskaNelemans19}, we include more recent works: \cite{Stefanon21} ($z$=5.5-10.5), \cite{Hamadouche24} ($z$=0.25-2.25), \cite{Navarro-Carrera24} ($z$=3.5-8.5) and \cite{Weibel24} ($z$=3.5-9.5). This more than doubles the number of GSMF estimates used at $z$>7.5, and leads to a more gradual evolution compared to \citetalias{ChruslinskaNelemans19}, where the GSMF shows a sharp drop between $z$=7-8. At lower $z$ the differences are minor. 
  Following  \citetalias{ChruslinskaNelemans19}, we distinguish between the (Schechter-like) high-mass part of the GSMF and its power-law low mass end.  
  At $\rm log_{10}(M_{*\text{}})\geqslant log_{10}(M_{lim})$, the GSMF is described as:
  \begin{equation}
    \label{eq: GSMF}
    \Phi \left(M_{*\text{}},z\right)=\ln(10)\,\Phi_*\left(\dfrac{M_{*\text{}}}{M_{\text{GSMF}}}\right)^{\alpha_{\text{high}}+1}\exp\left(-\dfrac{M_{*\text{}}}{M_{\text{GSMF}}}\right),
\end{equation}
where \begin{math}
    \Phi \left(M_{*\text{}},z\right)=\dfrac{dn_{\text{gal}}}{d\log_{10} M_{*\text{}}}
\end{math}
\\
Table \ref{app tab GSMF} summarizes the parameters of a single Schechter function fits to the high-mass part of the GSMF used in this work.
At $\rm log_{10}(M_{*\text{}})<log_{10}(M_{lim})$, the GSMF is described as a power-law with slope $\alpha_{\rm GSMF}$, i.e. \begin{math} \Phi\left(M_{*\text{}},z\right) \propto M_{*}^{\alpha_{GSMF}+1} \end{math}, 
  and is normalised to ensure continuity with the high mass part.  We assume $\rm log_{10}(M_{lim}/M_{\odot})= 7.7+0.4 z$ at $z<$4.5 and $\rm log_{10}(M_{lim}/M_{\odot})$=9.5 at $z\geq$4.5 and the slope $\alpha_{\rm GSMF}(z) = -1.4 -0.08 \cdot z$, reaching $\alpha_{\rm GSMF}$=-2.2 at $z=10$. Following \citetalias{ChruslinskaNelemans19}, we contrast this with the variation in which the slope is fixed to $\alpha_{\rm GSMF} = \alpha_{\rm fix}=-1.45$, corresponding to the average literature value between $z$=0 and $z$=2. 
 This serves to discuss the effect of low $M_{*}$ extrapolation on our results. 
 We extrapolate the GSMF down to $M_{*}=10^{6} \Msun$, where it is essentially unconstrained, making it one of the main sources of uncertainty in f$_{\rm SFR}$(Z,z) at $z\gtrsim$4 (\citetalias{ChruslinskaNelemans19,Chruslinska21}).
 The resulting GSMF is shown in the top panel of fig. \ref{fig: scaling rel.}.

\begin{table}[]
\caption{Parameters of equation \ref{eq: GSMF}, approximating the high-mass part of the GSMF.}
\centering
\resizebox{\columnwidth}{!}{
\begin{tabular}{cccc}
$z$ &
  log$_{10}$($\Phi_*$/$\rm Mpc^{-3} dex^{-1}$) &
  log$_{10}$(M$_{\rm GSMF}$/M$_{\odot}$) &
  $\alpha_{\rm high}$ \\ \hline
0   & -2.80 & 10.77 & -1.32  \\
1   & -3.02 & 10.79 & -1.34  \\
2   & -3.39 & 10.90 & -1.37  \\
3   & -4.23 & 11.15 & -1.66  \\
4   & -4.90   & 11.20 & -1.95 \\
5.5 & -5.60   & 11.15  & -2.25 \\
7   & -5.36 & 10.70   & -2.1   \\
8   & -5.60   & 10.50   & -1.98 \\
9   & -6.00     & 10.60   & -1.98 \\
10  & -6.40 & 10.43  & -1.99 \\
11  & -5.65 & 9.40  & -2.01
\end{tabular}
}
\label{app tab GSMF}
\end{table}

 We do not extend our calculations beyond $z=10.5$, except for extrapolating the cSFH trends shown in Fig. \ref{fig: cSFH}. For this purpose, we assume that at $z>10.5$ the high-mass GSMF is fixed and we continue with the $z$-dependent low-mass end as described above.
 We warn that such extrapolation would lead to an increase in the SFRD at $z>13$.

\subsection{Starburst sequence}\label{sec, app: SB}
\begin{figure}[]
    \centering
    \includegraphics[width=\columnwidth]{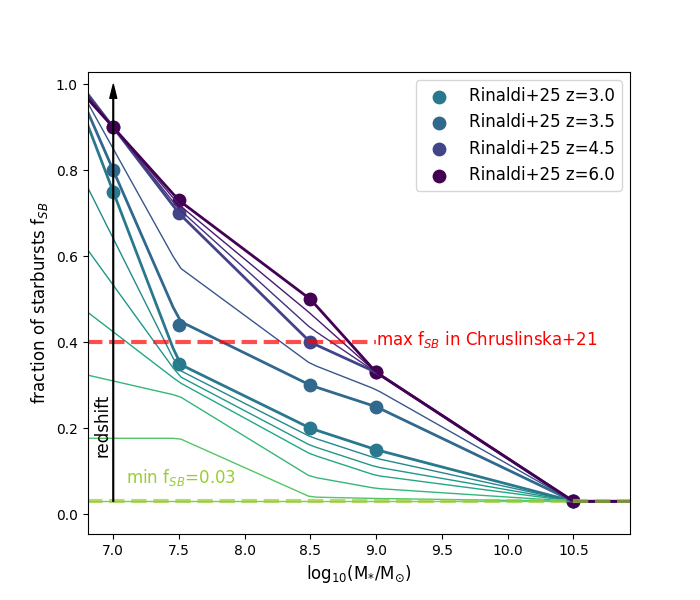}    
    \caption{
Fraction of starbursts as a function of galaxy stellar mass. Lines correspond to redshifts between 0 and 6 with a step of 0.5.
For $z<3$ and log$_{10}$(M$_{*}$/M$_{\odot}$)$>$8 the dependence is the same as in \citetalias{Chruslinska21}, here extended to $z>3$ and lower log$_{10}$(M$_{*}$) following the same approach but using the recent estimates from \cite{Rinaldi25} (circles, fig. 5 therein). 
}
    \label{fig: starburst frac}
\end{figure}

Following \citetalias{Chruslinska21}, in the negligible SB contribution variation we assume the SB sequence
$\rm log_{10}\mathrm{SFR}_{\mathrm{SB}}(M_\ast,z)=log_{10}\mathrm{SFR}_{\mathrm{SFMR}}(M_\ast,z) + \Delta_{SB}$ with $\Delta_{SB}$=0.59 dex, $\sigma_{\mathrm{SB}}=0.2$ dex and a fixed f$_{SB}$=0.03.
\\
The high-SB scenario from \citetalias{Chruslinska21} is based on
the interpolation between the results from \citet{Bisigello18} and \citet{Caputi17} and fixed at $z>4.4$ (exceeding $z$ of their samples). The resulting f$_{SB}$ increases to low M$_{*}$ and high $z$ up to f$_{SB}$=0.4, and the SB sequence is 1 dex above the SMFR. 
The work of \citet{Caputi17} has recently been extended to $z\sim 6$ and low M$_{*}\lesssim 10^{7}\Msun$ by \citet{Rinaldi22,Rinaldi25}.
They report  further increase in f$_{SB}$ with $z$ and decreasing M$_{*}$, and the lack of a clear distinction between the SB and MS at low log$_{10}$(M$_{*}$/M$_{\odot}$)$<$7.5. \citet{Rinaldi25} suggest that the majority of the cosmic SFRD is produced by the SB galaxies. We modify the f$_{SB}$(M$_{*}$, $z$) from \citetalias{Chruslinska21} to include the results from \citet{Rinaldi25} as shown in Fig. \ref{fig: starburst frac} to comment on the impact of such a high SB contribution on our results.
However, we assume the SB sequence is 1 dex above the SMFR.
This is smaller than the offset between the peaks in the sSFR distribution of SB and MS galaxies in \citet{Rinaldi25} (see Fig. 3 therein). Additionally, for variations with $a_{\rm SFR}$<1 (which is the case for \citealt{Rinaldi22,Rinaldi25}), we require this offset to decrease linearly from 1 dex to 0 dex between log$_{10}$(M$_{*}$/M$_{\odot}$)=9 and log$_{10}$(M$_{*}$/M$_{\odot}$)=6. 
This is necessary to avoid drastically overestimating the cosmic SFRD($z>2$) (right panel in fig. \ref{fig: cSFH}, thick gray lines).\\
We note that in terms of the results presented in this paper, the high $f_{SB}$ scenario described above would be indistinguishable from a scenario in which the SB sequence is absent and a$_{SFR}$ is $z$-dependent (a$_{SFR}\sim$0.9 at $z=0$ and becomes a$_{SFR}\lesssim$0.8 at $z>2$).

\section{Overview of model variations}\label{app: variations}
Table \ref{tab: variations} summarizes the variations of assumptions used to describe the statistical properties of the galaxies needed to derive $f_{\rm SFR}(Z_{O/H},t)$ and $f_{\rm SFR}(Z_{Fe/H},t)$ that were considered in this study.

\begin{table*}[]
\centering
\caption{Overview of the assumptions (model variations) considered in this study. Combination of assumptions highlighted in bold corresponds to our example variation.}

\resizebox{\textwidth}{!}{%
\begin{tabular}{ll}
Variation                & Notes                                                                                                 \\ \hline 
\multicolumn{2}{c}{[O/Fe] -- sSFR relation (following \citetalias{Chruslinska24_OFe},  see sec. \ref{sec: which O/Fe-sSFR})}                                                                                      \\  \hline 
"fast" Fe enrichment      & $f_{\rm Ia}\propto t^{-1}$, $\tau_{\rm Ia; min}$=40 Myr, m$^{CCSN}_{Fe}=0.03 M_{\odot}$              \\
\textbf{"mixed" Fe enrichment}     & $f_{\rm Ia}$ following \citet{Greggio10}, m$^{CCSN}_{Fe}=0.05 M_{\odot}$                                       \\
"slow" Fe enrichment      & $f_{\rm Ia}\propto t^{-1}$, $\tau_{\rm Ia; min}$=400 Myr, m$^{CCSN}_{Fe}=0.1 M_{\odot}$              \\  \hline 
\multicolumn{2}{c}{SFMR/ galaxy main sequence}                                                                                   \\ \hline 
a$_{\rm SFR}$=1           & following eq. 15 and table 2 in \citetalias{Popesso23}                                               \\
a$_{\rm SFR}$=0.8         & \citetalias{Popesso23} with shallower slope: parameter a4 in Tab. 2 in \citetalias{Popesso23} set to 0.8                         \\
\textbf{a$_{\rm SFR}$=1 + evol.} &
  \citetalias{Popesso23} with faster evolution at $z>1.8$: at $z\geqslant$1.8 parameter a1 in Tab. 2 in \citetalias{Popesso23} set to -0.3 \\  \hline 
\multicolumn{2}{c}{GSMF (following \citetalias{ChruslinskaNelemans19} with updates, see appendix \ref{sec, app: GSMF})}                                                                                                         \\  \hline 
\textbf{$\alpha_{\rm GSMF}(z)$}   & low mass end slope steepening with $z$: $\alpha_{\rm GSMF}= -1.4 -0.08 \cdot z$                                                              \\
$\alpha_{\rm GSMF}$ fixed & $\alpha_{\rm GSMF}$=-1.45                                                                            \\  \hline 
\multicolumn{2}{c}{MZR($z=0$) normalisation }                                                                                \\  \hline 
\textbf{$Z_{\rm O/H; MZR0}$ = 9}   & fiducial value used here, following \citet{Curti20} but shifted by +0.2 dex to match recombination line based estimates \\
$Z_{\rm O/H; MZR0}$=8.8 - 9.1                 & range considered in \citetalias{Chruslinska21}, where relevant, we indicate how this would lead to a systematic shift in our results    \\  \hline 
\multicolumn{2}{c}{MZR(z=0)}                                                                                                     \\  \hline 
 - &
  eq. 7 in \citetalias{Chruslinska21} with a$_{\rm MZR}$=0.28, $\beta_{\rm MZR}$=0.23, $M_{\rm 0;MZR0}=10.02$ to match \citet{Curti20}, $Z_{\rm O/H; MZR0}$ as above \\  \hline 
\multicolumn{2}{c}{FMR($z>3$)}                                                                                                   \\  \hline 
fixed                     & redshift-invariant FMR modelled following \citetalias{Chruslinska21}                                 \\
\textbf{evol.} &
  evolving normalisation at $z>$3, parameter $Z_{\rm O/H; 0}$ in eq. 1 in \citetalias{Chruslinska21} set to $Z_{\rm O/H; 0}(z>3)$ = -0.0357 ($z$ - 3) + $Z_{\rm O/H; MZR0}$ \\  \hline 
\multicolumn{2}{c}{contribution of starbursts}                                                                                   \\  \hline 
\textbf{negligible}             & following \citetalias{Chruslinska21}, fixed SB fraction $f_{SB}$=0.03, SB sequence 0.59 dex above the SFMR  and scatter $\sigma_{SB}$=0.2 dex    \\
high / high $f_{SB}$      & see appendix \ref{sec, app: SB}, based on \citet{Rinaldi25}  and considered only  with $a_{\rm SFR}$=1 SFMR variation                                        \\
\end{tabular}%
}
\label{tab: variations}
\end{table*}

\section{Metallicity spread in galaxies}\label{app: scatter}
\begin{figure}[htbp]
    \centering
    \includegraphics[width=0.8\columnwidth]{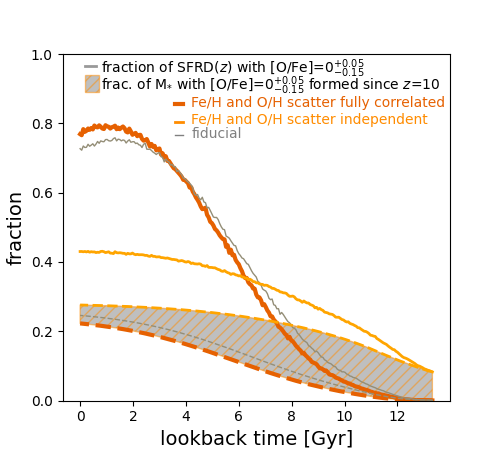}
    
    \caption{
Effect of the assumed [O/Fe] scatter (relative to the fiducial case) on the estimated fraction of the SFRD forming with near-solar [O/Fe] (solid lines) and on the cumulative fraction of stellar mass formed with near-solar [O/Fe] since $z=10$ (dashed lines and gray area). Larger scatter (the fully independent $Z_{\rm O/H}$ and $Z_{\rm Fe/H}$ case) increases these fractions, while smaller scatter (the fully correlated $Z_{\rm O/H}$ and $Z_{\rm Fe/H}$ case) decreases them.
    }
    \label{fig: near_solar_scatter_effect}
\end{figure}

The SFR within each "galaxy" in our framework is distributed over a range of metallicities. 
As in \citetalias{ChruslinskaNelemans19}, we model this as a normal distribution centered on the oxygen abundance $Z_{\rm O/H;FMR}$ assigned to a "galaxy" based on the FMR (given SFR, $M_{\ast}$, and $z$), with dispersion $\sigma_{\nabla \rm O/H}=0.14$ dex. This is a major simplification, but it does not significantly affect cosmic–volume–averaged results (e.g., metallicity–dependent cosmic SFH). \citetalias{ChruslinskaNelemans19} motivate this choice with the distribution of HII–region metallicities in $z\sim0$ disk galaxies from CALIFA \citep{Sanchez-Menguiano16,Sanchez-Menguiano17}, which is roughly symmetric about the average value at the effective radius $R_{e}$; we take that average to represent the $Z_{\rm O/H}$ assigned via the scaling relations.
\\
Here we further assume that $Z_{\rm Fe/H}$ within each "galaxy" follows an analogous normal distribution with the same dispersion, $\sigma_{\nabla \rm Fe/H}=\sigma_{\nabla \rm O/H}$, centered on the value  set by the [O/Fe]–sSFR relation and $Z_{\rm O/H;FMR}$ (eq. \ref{eq: avg Fe/H}). This additional simplification is justified by the broadly comparable $Z_{\rm O/H}$ and $Z_{\rm Fe/H}$ gradients observed in galaxies \citep[][]{Hernandez21}.
\\
One must still specify the spread in [O/Fe], $\sigma_{\rm O/Fe}$. If $Z_{\rm O/H}$ and $Z_{\rm Fe/H}$ with which the stars form inside a galaxy were fully correlated, then $\sigma_{\rm O/Fe}=0$ and, at a given $z$, all of the SFR within a "galaxy" would share a single [O/Fe]. 
If instead the two were uncorrelated (e.g., in a highly turbulent ISM mixing), then [O/Fe] would follow a normal distribution with $\sigma_{\rm O/Fe}=\sqrt{2} \times \sigma_{\nabla \rm O/H}$. Reality likely lies between these limits; for example, MW stars at fixed age, location, and [Fe/H] show a moderate spread in [O/H]. As our fiducial choice we adopt a weakly uncorrelated case, $\sigma_{\rm O/Fe}=\sigma_{\nabla \rm O/H}/4$. This choice is arbitrary, but it affects only $f_{\rm SFR}(\mathrm{[O/Fe]},z)$ (see eq. \ref{eq:notes_full_csfh_ofe}) and results derived from it (Fig. \ref{fig: OFe_avg}, Fig. \ref{fig: f_SFR_OFe_t_mixed}, Fig. \ref{fig: f_SFR_OFe_t_fast}).
\\
Figure \ref{fig: near_solar_scatter_effect} illustrates the impact of $\sigma_{\rm O/Fe}$ on the fraction of the cosmic SFRD and the integrated stellar mass formed with near-solar [O/Fe] (as in the top panel of Fig. \ref{fig: OFe_avg}). The latter quantity is computed by evaluating Eq.~\ref{eq:notes_near_solar_ofe_mass}:
\begin{equation}
\begin{split}
\frac{
 \int_{t(z=10)}^{t} \int_{-0.15 + \mathrm{Z_{O/Fe}}}^{ 0.05 + \mathrm{Z_{O/Fe}}_{\odot}} f_{\mathrm{SFR}}(\mathrm{Z_{O/Fe}},t') \ dZ_{O/Fe}\ dt'}{\int_{t(z=10)}^{t} \int_{Z_{O/Fe;min}}^{Z_{O/Fe;max}} f_{\mathrm{SFR}}(\mathrm{Z_{O/Fe}},t') \ dZ_{O/Fe}\ dt'}
\end{split}
\label{eq:notes_near_solar_ofe_mass}
\end{equation}
where $[\mathrm{O/Fe}] = Z_{O/Fe} - Z_{O/Fe,\odot}$. The integration limits in the numerator span oxygen-to-iron abundance ratios encompassing the range of solar values adopted in the literature, while the denominator covers the full range of galaxy $Z_{O/Fe}$ values. The cosmic time integral extends from the epoch corresponding to $z=10$ (the maximum redshift considered in this work) to time $t$.
The gray band in the top panel of Fig.~\ref{fig: OFe_avg} is obtained by evaluating Eq.~\ref{eq:notes_near_solar_ofe_mass} across our model variations (Table~\ref{tab: variations}) assuming $\sigma_{\rm O/Fe}=\sigma_{\nabla \rm O/H}/4$ and it spans the full range of resulting values.
In general, larger $\sigma_{\rm O/Fe}$ broadens the distribution of $[\mathrm{O/Fe}]$ at a given epoch, increasing the fraction of stellar mass formed with near-solar $[\mathrm{O/Fe}]$ and leading to a weaker evolution of the instantaneous near-solar SFRD fraction over cosmic time.
In Fig.~\ref{fig: near_solar_scatter_effect} we illustrate this effect for the model variation that maximises the near-solar $[\mathrm{O/Fe}]$ fraction (i.e. it sets the upper edge of the gray band in Fig. \ref{fig: OFe_avg}), obtaining a conservative upper bound of $\lesssim 30\%$ on the fraction of stellar mass formed with near-solar $[\mathrm{O/Fe}]$ (yellow dashed line in Fig. \ref{fig: near_solar_scatter_effect}, obtained for $\sigma_{\rm O/Fe}=\sqrt{2} \cdot \sigma_{\nabla \rm O/H}$, corresponding to independent scatter in oxygen and iron). We therefore conclude that the specific treatment of scatter does not affect our main result that star formation with near-solar $[\mathrm{O/Fe}]$ is rare.

\section{LGRB hosts: HI-weighted mean metallicity in redshift bins}

Table \ref{tab:LGRB binned} summarizes the HI column density-weighted mean metallicity in redshift bins for the sample of LGRB hosts discussed in  Section \ref{sec: LGRB sample}.

\begin{table}[]
\caption{$N_{\rm HI}$-weighted mean metallicity 
with $z$ based on absorption-derived LGRB host properties.}
\centering
\resizebox{0.8\columnwidth}{!}{%
\begin{tabular}{ccccc}
\multicolumn{5}{c}{selected subsample for which [M/H]$_{\rm tot} \approx$ [Fe/H]}                       \\
$z_{\rm low}$ & $z_{\rm high}$ & number in bin & $\langle \mathrm{[Fe/H]} \rangle_{\rm HI}$ & dispersion \\ \hline
1.7  & 2.1  & 6  & -0.65 & 0.38 \\
2.1  & 2.5  & 6  & -1.01 & 0.46 \\
2.5  & 2.95 & 5  & -1.2  & 0.51 \\
2.95 & 6.4  & 5  & -1.69 & 0.23 \\ \hline
\multicolumn{5}{c}{all LGRBs}   \\ 
$z_{\rm low}$ & $z_{\rm high}$ & number in bin & $\langle \mathrm{[M/H]} \rangle_{\rm HI}$ & dispersion \\ \hline
1.7  & 2.2  & 9  & -0.72 & 0.64 \\
2.2  & 2.8  & 10 & -1.01 & 0.47 \\
2.8  & 3.8  & 9  & -1.21 & 0.34 \\
3.8  & 6.4  & 9  & -1.18 & 0.68 \\ \hline
\end{tabular}%
}
\tablebib{See Section \ref{sec: LGRB sample} and  Fig. \ref{fig: lgrb combined}.
Redshift bin edges (column 1 and 2) were selected to ensure that each bin contains a similar number of LGRB hosts.
Top rows: subsample with dust corrections derived from Zn or accounting for $\alpha$-enhancement for which [M/H]$_{tot}\approx$[Fe/H].
Bottom rows: also including objects with more approximate dust correction based on $\alpha$-elements, in which case [M/H]$_{tot}$ may overestimate [Fe/H].
}
\label{tab:LGRB binned}
\end{table}

\section{$Z_{\rm O/H}$ and $Z_{\rm Fe/H}$ metallicity distribution functions} \label{sec: MDFs}

\begin{figure*}[htbp]
    \centering
    \includegraphics[width=0.8\textwidth]{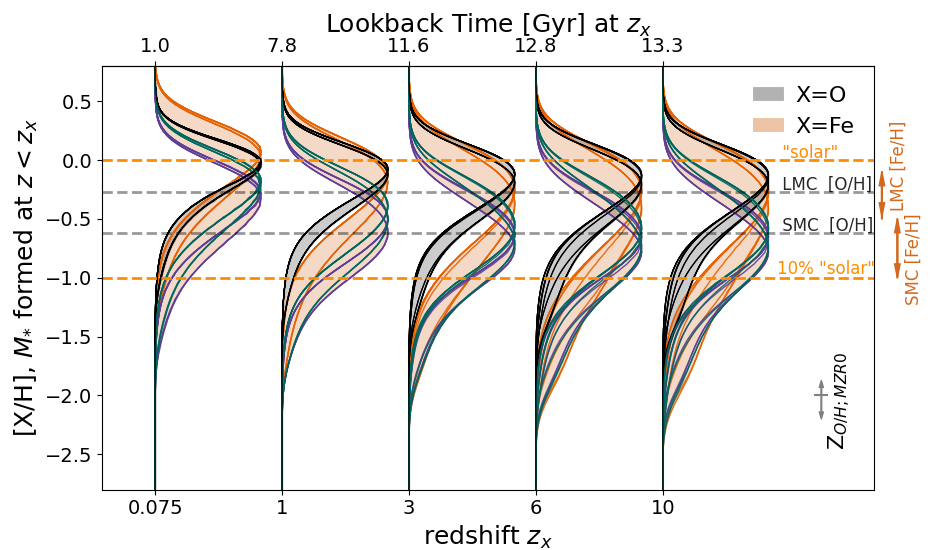}
    
    \caption{
    Metallicity distribution function of stellar mass formed between $z=0$ and $z_{x}$ (as indicated on the horizontal axis), for [O/H] (gray bands, black curves) and [Fe/H] (brown bands, coloured curves; colours distinguish the [O/Fe]–sSFR relation as in Fig. \ref{fig: OFe-sSFR rel.}). Bands show the range spanned by individual model variations, which are shown as lines. Dashed horizontal lines mark solar, 0.1 solar, and LMC/SMC [O/H] (\citealt[][]{GrevesseSauval98} reference scale); LMC/SMC [Fe/H] are marked as arrows. Grey arrows indicate the systematic uncertainty (not accounted for in the plotted ranges) from the $Z_{\rm O/H}$ calibration, which shifts all curves by a constant factor.
    All MDFs are normalised to unity to compare the shapes.
    }
    \label{fig: z_integral_low_to_high}
\end{figure*}

\begin{figure*}[htbp]
    \centering
    \includegraphics[width=0.8\textwidth]{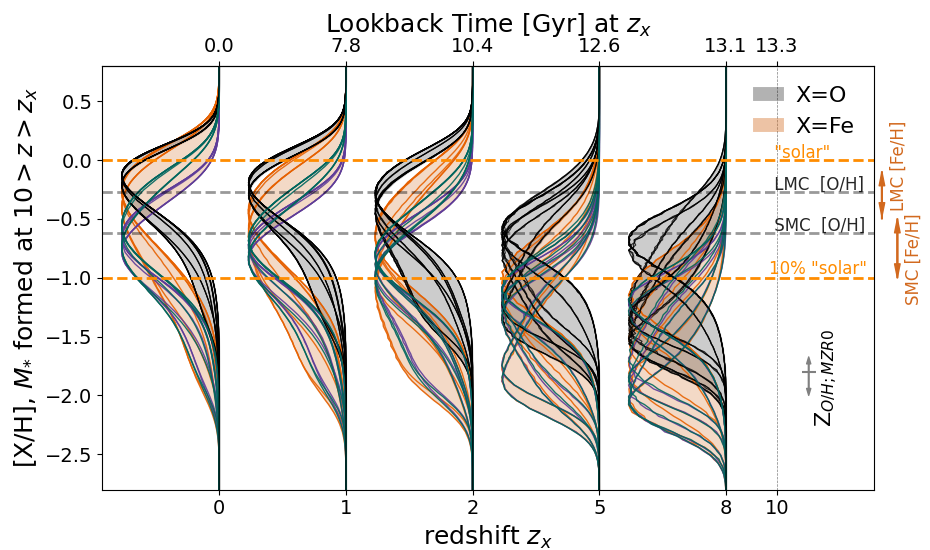}
    
    \caption{
    Metallicity distribution function of stellar mass formed between $z_{x}$ (as indicated on the horizontal axis) and $z=10$, for [O/H] (gray bands, black curves) and [Fe/H] (brown bands, coloured curves; colours distinguish the [O/Fe]–sSFR relation as in Fig. \ref{fig: OFe-sSFR rel.}). Bands show the range spanned by individual model variations, which are shown as lines. Dashed horizontal lines mark solar, 0.1 solar, and LMC/SMC [O/H] (reference scale \citep[][]{GrevesseSauval98}); LMC/SMC [Fe/H] are marked as arrows. Gray arrows indicate systematic shifts from $Z_{\rm O/H}$ calibration.
    }
    \label{fig: z_integral_high_to_low}
\end{figure*}

In this section we compare the $f_{\rm SFR}(Z_{\rm Fe/H},z)$ and $f_{\rm SFR}(Z_{\rm O/H},z)$ integrated over $z$ - i.e. the $Z_{\rm Fe/H}$- and $Z_{\rm O/H}$-metallicity distribution functions (MDFs).
Figure \ref{fig: z_integral_low_to_high} shows the MDFs of stars formed between $z=0$ and increasing $z_{x}$: the leftmost curves correspond to stars formed in the last $\sim$1 Gyr, while the rightmost curves encompass the full $\approx$13.3 Gyr of cosmic history. For comparison, Fig. \ref{fig: z_integral_high_to_low} presents the complementary case, showing MDFs of stars formed between $z=10$ and decreasing $z_{x}$, i.e. focusing on the early Universe.
Together, these figures illustrate the buildup of the cosmic MDF, with the peak established at $z \lesssim 3$, where most of the cosmic stellar mass forms, and the low-metallicity tail developed at $z \gtrsim 2$ (cf. Fig.~\ref{fig: z_integral_high_to_low}).
The brown/black ranges in  Fig. \ref{fig: z_integral_low_to_high} span between the model variations and broaden with increasing $z_{x}$, reflecting the fact that our estimates are more uncertain at high $z$.
As before, different assumptions about the [O/Fe]–sSFR relation (which affect only the $Z_{\rm Fe/H}$-MDFs) are distinguished by colour.
For stars formed in the last $\sim$1 Gyr, this assumption shifts
the peak of the $Z_{\rm Fe/H}$-MDF to higher (orange lines, 'fast' Fe enrichment) or lower (green/purple lines, 'mixed'/'slow' Fe enrichment) metallicities relative to solar and to the $Z_{\rm O/H}$-MDF peak. The offset between the peaks reaches up to $\sim$0.3 dex for MDFs integrated over the past 1 Gyr and up to $\gtrsim$0.5 dex when integrating over the full $z$ range spanned by our models (with 'slow' Fe enrichment maximizing the difference). For any fixed variation, the $Z_{\rm Fe/H}$-MDF is broader than the corresponding $Z_{\rm O/H}$-MDF. Aside from the impact of the [O/Fe]–sSFR relation (i.e., comparing curves of the same colour), differences appear mainly in the low-metallicity tails.
This regime is dominated by contributions from low-$M_{*}$ galaxies, where the associated uncertainties (parametrized by $a_{\rm SFR}$ and $\alpha_{\rm GSMF}$) accumulate in our model.
We caution that all MDFs in Figs.~\ref{fig: z_integral_high_to_low} and \ref{fig: z_integral_low_to_high} are normalized to unity to facilitate comparison of their shapes, but the total stellar mass formed differs across models (see Sec.~\ref{sec: low-Z cSFH}). In particular, variations producing more extended low-metallicity tails generally also yield higher SFRDs (by a factor of $\sim$2–3 at $z<2$, and increasing to higher $z$), due to the larger contribution of low-$M_{*}$ galaxies.

\section{Additional figures}

\begin{figure}[]
    \centering
    \includegraphics[width=\columnwidth]{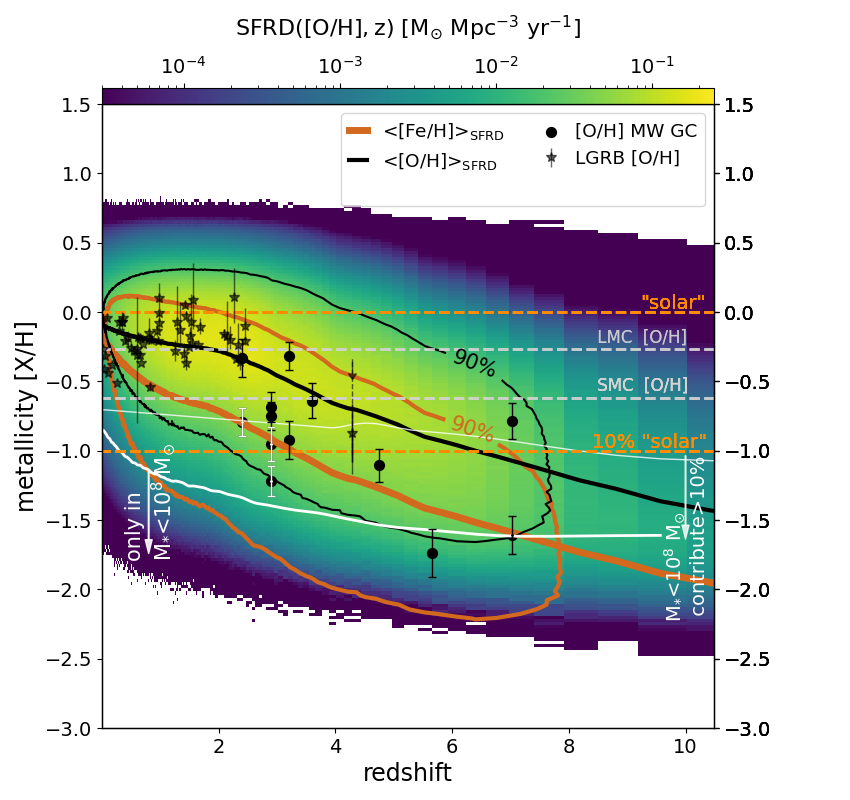}    
    \caption{
Distribution of the cosmic SFRD at different [O/H] and $z$ (colour scale), shown for our example variation.
Black/brown lines: SFRD-weighted average [O/H] / [Fe/H] at each $z$.
White solid lines: [O/H] below  which  $M_{*}<10^{8}\Msun$ galaxies contribute at least 10\% of the SFRD (thin line) and below which contribution of $M_{*}>10^{8}\Msun$ galaxies is negligible (thick line). 
Stars: [O/H] of individual LGRB hosts \citep{Graham23}.
Dots: [O/H] of MW globular clusters. 
[O/H] of the LMC and SMC on our solar scale are indicated for the reference. 
}
    \label{fig: SFRD_OH_z_example}
\end{figure}

\begin{figure}[htbp]
    \centering
    \includegraphics[width=\columnwidth]{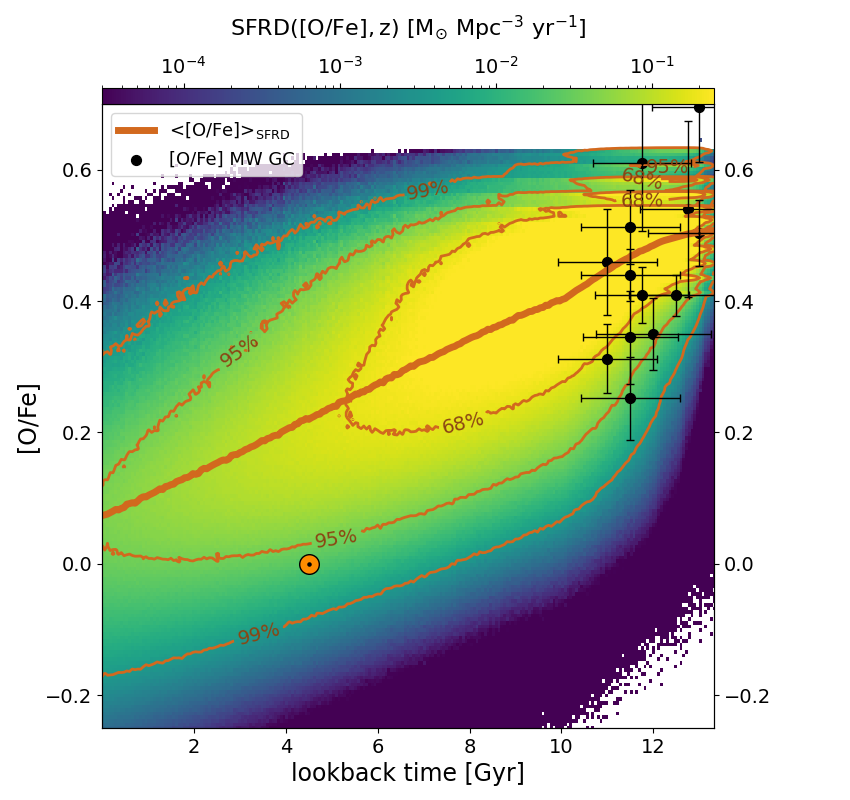}
    
    \caption{
Distribution of the cosmic SFRD at different [O/Fe] and lookback time (colour scale), shown for our example variation, in particular, assuming the [O/Fe]-sSFR relation corresponding to our "mixed Fe enrichment" case.
Brown line: SFRD-weighted average [O/Fe] at each $z$.
Contours enclose 68\%, 95\% and 99\% of the SFRD.
Black dots: [O/Fe] of MW globular clusters. 
Orange dot marks the Sun's formation epoch at [O/Fe]=0 assuming solar scale from \citet{GrevesseSauval98}.
 }
    \label{fig: f_SFR_OFe_t_mixed}
\end{figure}

\begin{figure}[htbp]
    \centering
    \includegraphics[width=\columnwidth]{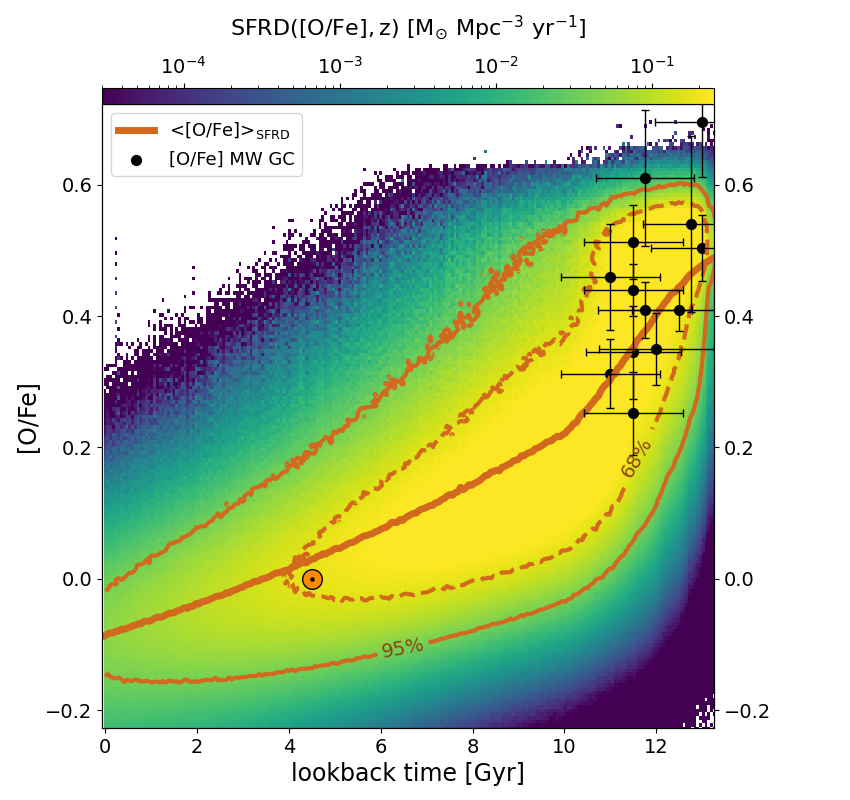}
    
    \caption{
Same as Fig. \ref{fig: f_SFR_OFe_t_mixed}, but  assuming the [O/Fe]-sSFR relation corresponding to our "fast Fe enrichment" case.
    }
    \label{fig: f_SFR_OFe_t_fast}
\end{figure}

Figures \ref{fig: SFRD_OH_z_example} and \ref{fig: f_SFR_OFe_t_mixed} show the $f_{\rm SFR}([O/H], z)$ and $f_{\rm SFR}([O/Fe], t)$ distributions for our example variation, respectively. Figure \ref{fig: f_SFR_OFe_t_fast} shows the $f_{\rm SFR}([O/Fe], t)$ for the same variation but assuming a [O/Fe]-sSFR relation corresponding to our "fast Fe enrichment" case.
Figure \ref{fig: z_integral_high_to_low} shows the MDF of stellar mass formed between decreasing $z$ thresholds and $z=10$, i.e. complementary to Fig. \ref{fig: z_integral_low_to_high}.

\end{appendix}

\end{document}